\documentclass[%
 reprint,
 amsmath,amssymb,dff
 aps,article,
 pra,
]{revtex4-1}

\usepackage{graphicx}
\usepackage{epstopdf}
\usepackage{epsfig}
\usepackage{caption}
\usepackage{subcaption}
\usepackage{dcolumn}
\usepackage{bm}
\usepackage{physics}
\usepackage{amsthm}
\usepackage[utf8]{inputenc}
\usepackage[english]{babel}
\usepackage{etoolbox}
\usepackage{color}
\begin{document}

\title{Amplifying hybrid entangled states and superpositions of coherent states}

\author{InU Jeon}
\author{Sungjoo Cho}
\author{Hyunseok Jeong}
\email{h.jeong37@gmail.com}
\affiliation{%
IRC NextQuantum, Department of Physics \& Astronomy, Seoul National University, Seoul 08826, Korea}

\date{\today}

\begin{abstract}
We compare two amplification schemes, photon addition and then subtraction ($\hat{a}\hat{a}^\dagger$) and successive photon addition ($\hat{a}^\dagger{}^2$), applied to hybrid entangled states (HESs) and superpositions of coherent states (SCSs). We show that the fidelity and gain of the amplification schemes for HESs are the same as those of coherent states. However, SCSs show quite non-trivial behavior by the amplification schemes, depending on the amplitudes of coherent states, the number of coherent-state components, and the relative phases between the components. This implies that appropriate amplification schemes for SCSs should be chosen depending on the tasks and specific forms of the states. To investigate the quality of amplified states, we calculate the quantum Fisher information, a measure of quantum phase estimation. In terms of the quantum Fisher information, the $\hat{a}\hat{a}^\dagger$ scheme tends to show better performance for relatively small amplitudes ($\alpha\lesssim 0.9$), while the $\hat{a}^\dagger{}^2$ scheme is better in the larger amplitude regime. The performance of the two schemes becomes indistinguishable as the amplitude grows sufficiently large.
\end{abstract}

\maketitle

\section{introduction}

Hybrid entangled states (HESs)~\cite{Jeong14,Morin14} 
and superpositions of coherent states (SCSs)~\cite{Yurke86, Schleich91} in free-traveling fields are well-known nonclassical states.
A HES is an entangled state between a discrete-variable (DV)  system such as a photon number state and a continuous-variable (CV) system such as a coherent state. This integrated perspective toward the traditional dichotomy of CV and DV quantum information brings new advantages that could not have been obtained in each area individually. Apart from its theoretical versatility, interplay, and conversion between CV and DV systems~\cite{Jeong23} have been experimentally realized~\cite{Ulanov17, Sychev18, Darras23} indicating generation and processing of hybrid entanglement are well within our reach. HESs are also useful for fundamental studies on quantum physics such as loss-tolerant verification of Bell-inequality violations~\cite{Kwon13}, quantum teleportation~\cite{Park12, Lee13, Jeong16, Kim16, Ulanov17, Sychev18}, quantum computation~\cite{Lee13, Omkar19}, quantum communication~\cite{Bose23, Lim16, JK2001}  and steering verification~\cite{Cavailles18, Huang19}. Generation schemes for HESs were theoretically proposed~\cite{Gerry99, Nemoto04, Jeong05a, Anderson13,Jeong14,Morin14} and experimentally implemented~\cite{Jeong14, Morin14}.  SCSs are superpositions of coherent states with different phases, and a typical example is a superposition of two coherent states with opposite phases. They are useful for many quantum information tasks such as quantum computation~\cite{Cochrane99, Jeong02, Ralph03, Lund08, Myers11, Kim15}, quantum teleportation~\cite{vanEnk01, Jeong01, Neergaard-Nielsen13}, precision measurements~\cite{Gerry01, Gerry02, Ralph02, Munro02, Campos03, Joo11, Hirota11, Joo12, Zhang13}, non-Gaussian state generation ~\cite{Vasconcelos10, Weigand18, Shunya24} and error suppression~\cite{Li17, Hastrup22}.
They have also been used to investigate violations of local realism \cite{Wilson02, Park15} and nonlocal realism \cite{Paternostro10,Lee11}.
SCSs in free-traveling fields have been implemented approximately using various methods~\cite{Ourjoumtsev06, Neergaard-Nielsen06, Ourjoumtsev07,  Wakui07, Ourjoumtsev09, Jeong05}
In these regards, obtaining sufficiently large amplitudes of continuous variables in HESs or SCSs is required to accomplish various quantum information processing tasks.

One way to obtain HESs or SCSs of large amplitudes is to amplify the states~\cite{Ralph03, Lund04, Neergaard-Nielsen13, Lag13, Lvovsky2017, Oh18} or purify them~\cite{JK2001, Sheng13}.
Although deterministic and noiseless linear amplification for a bosonic system is forbidden by quantum mechanics~\cite{Caves88}, there have been studies on non-deterministic amplification of coherent states \cite{Parigi07,Fadrny24,Zavatta11,Park16}.
In this paper, we adopt the photon addition and then subtraction scheme~\cite{Parigi07, Zavatta11} ($\hat{a}\hat{a}^\dagger$) and the successive photon addition scheme~\cite{Park16,Fadrny24} ($\hat{a}^\dagger{}^2$). We compare the fidelity and gain of the two schemes and take the quantum Fisher information {\it for phase estimation} ~\cite{Helstrom76, Braunstein94, Barndorff-Nielsen00, Paris09, Petz10} as an indicator of amplification performance.
This could be a practical measure to estimate the amplification performance because various non-classical states of light fields are utilized to estimate the quantum phase~\cite{Usuga10, Genoni11, Berni15, Izumi16}.
In fact, in a previous work~\cite{Park16}, the quantum Fisher information is used as a measure for amplification schemes using $\hat{a}\hat{a}^\dagger$ and $\hat{a}^\dagger{}^2$ for $2$-dimensional SCSs. 
%
Extending SCSs to multidimensional cases would be consistent with employing the same measure to show the performance of amplified states. However, we do not adopt the equivalent-input-noise (EIN)~\cite{Grangier92} as a measure of noiseless amplification because the expectation value of a quadrature operator, $\langle x_\lambda \rangle$, is zero for all $\lambda$ for HESs and SCSs.

Consequently, we show that the $\hat{a}\hat{a}^\dagger$ scheme gives higher fidelity while the $\hat{a}^\dagger{}^2$ scheme gives higher gain. Performance of optimal quantum phase estimation measured by the quantum Fisher information shows that the $\hat{a}\hat{a}^\dagger$ scheme is better for relatively small amplitudes, where the exact threshold value differs with respect to quantum states. The $\hat{a}^\dagger{}^2$ scheme is better for larger amplitudes, and the difference between their performances asymptotically vanishes as the input-state amplitude increases.

In Sec. II, we begin with preliminaries of HESs and SCSs. In Sec. III, we investigate and compare the two amplification schemes applied to HESs, and in Sec. IV, we investigate and compare those schemes applied to SCSs. Analyses of the success probabilities of amplification schemes acting on HESs and SCSs are provided in Sec. V. In Sec. VI, we conclude our paper by comparing the two amplification schemes.

\section{Preliminary}
\subsection{Hybrid entangled states}
The HESs are bosonic states consisting of entanglement between CV states and DV states. We define a qubit basis of HESs here as
 
  \begin{align} \label{2d hybrid}
        \begin{split}
            \ket{\mathcal{H}^0_\alpha} = \frac{1}{\sqrt{2}}(\ket{0,\alpha}+\ket{1,-\alpha}), \\      \ket{\mathcal{H}^1_\alpha} = \frac{1}{\sqrt{2}}(\ket{0,\alpha}-\ket{1,-\alpha}),
        \end{split}
  \end{align}
 where $\ket{n,\alpha}=\ket{n}\otimes\ket{\alpha}$, a product of a number state with photon number $n$ and a coherent state with the amplitude $\alpha$($\in\mathbb{R}^+$ for convenience). For a 2D HES system, The DV part can also be polarization states $\ket{H}$ and $\ket{V}$. Although coherent states $|\pm\alpha\rangle$ are not orthogonal to each other, orthogonality of number states provides independence of normalization factor with respect to the amplitude $\alpha$.
 
HES qubits can be generalized to represent qudits using d-fold complex roots of 1, $\omega_d := e^{\frac{2\pi i}{d}}$,
   \begin{align}
       \ket{\mathcal{H}^k_{\alpha,d}} := \frac{1}{\sqrt{d}}\sum_{n=0}^{d-1} \omega_d^{-kn}\ket{n, \alpha \omega_d^n},
   \end{align}
where $0\leq k<d$ and $k$ is a non-negative integer index for the $d$-dimensional qudit basis states $\ket{\mathcal{H}^k_{\alpha,d}}$. The overlap between arbitrary basis states can be calculated as
   \begin{align}\nonumber \label{hybrid orthogonality}
       \bra{\mathcal{H}^k_{\alpha,d}} \ket{\mathcal{H}^l_{\alpha,d}} &= \frac{1}{d} \sum_{m,n} \omega_d^{km} \omega_d^{-ln} \bra{m, \alpha \omega_d^m} \ket{n, \alpha \omega_d^n} \\
       &= \frac{1}{d} \sum_{n} \omega_d^{(k-l)n} = \delta_{k,l},
   \end{align}
which indeed shows orthonormality. Also, the photon number statistics of HESs are the same with coherent states with the same amplitude. We do not explicitly prove this since the statement can be directly derived from Proposition 1 in Sec. III.

Another important feature of the HES qudits is that photon addition or subtraction maps them to other HES qudits. For a $d$-dimensional HES $\ket{\mathcal{H}^k_{\alpha,d}}$, $m$ successive photon addition maps $\ket{\mathcal{H}^k_{\alpha,d}}$ to $\ket{\mathcal{H}^{l}_{\beta,d}}$ for some different amplitude $\beta$ where $l \equiv k+m(\mathrm{mod}\ d)$.
   \begin{widetext}
        \begin{align} \nonumber \label{Hybrid m successive}
             \frac{1}{\mathcal{N}}\bra{\mathcal{H}^l_{\beta,d}} (I \otimes (\hat{a}^\dagger)^m) \ket{\mathcal{H}^k_{\alpha,d}} = \frac{1}{\mathcal{N}d}\sum_n \omega_d^{(l-k)n} \bra{\beta \omega_d^n} (\hat{a}^\dagger)^m \ket{\alpha \omega_d^n}            &= \frac{\beta^m}{\mathcal{N}} \mathrm{exp}[-\frac{1}{2}(\alpha-\beta)^2] \sum_n \frac{1}{d} \omega_d^{(l-k-m)n}  \\
             &= \frac{1}{\mathcal{N}}\delta_{l,k+m ; d} \, \beta^m \mathrm{exp}[-\frac{1}{2}(\alpha-\beta)^2],
        \end{align}
    \end{widetext}
where $\delta_{m,k;d} := \delta_{m,k+d\cdot j}$ with $j\in\mathbb{Z}$. That is, $\delta_{m,k;d}=1$ if and only if $m \equiv k(\mathrm{mod}\ d)$. Also, $\mathcal{N} = \mathcal{N}(\alpha) = \sqrt{\sum_{k=0}^n \frac{(n!)^2}{k!((n-k)!)^2}\alpha^{2(n-k)}}$ is a normalization factor from photon additions. The normalization factor $\mathcal{N} \approx \alpha^n$ in the asymptotic limit of $\alpha \gg 1$. In such a limit the optimal choice of $\beta$ yielding maximal overlap becomes $\alpha$, and in that case, fidelity is strictly 1. Even if we consider a more realistic regime with smaller amplitudes($\alpha < 9$), we can achieve $F > 0.99$ with well-chosen $\beta$ up to $\approx$12 successive photon additions.

Successive photon subtraction acts more trivially to HES qudits than the photon addition. $m$ successive photon subtraction exactly maps $\ket{\mathcal{H}^k_{\alpha,d}}$ to $\ket{\mathcal{H}^{l'}_{\alpha,d}}$ with $l' \equiv k-m(\mathrm{mod}\ d)$. Note that we always observe the exact transition to target HES qudit without any assumption or change of amplitude.

\subsection{Superposition of coherent states}
As the name suggests, the superposition of coherent states (SCS) is a linear combination of coherent states with distinct phases. Typical SCS qubit basis, also known as even/odd cat states are given by
    \begin{align} \label{2d SCS}
        \begin{split}
            \ket{\mathcal{C}^0_\alpha} &= \frac{1}{\sqrt{2(1+e^{-2\alpha^2})}}(\ket{\alpha}+\ket{-\alpha}), \\
            \ket{\mathcal{C}^1_\alpha} &= \frac{1}{\sqrt{2(1-e^{-2\alpha^2})}}(\ket{\alpha}-\ket{-\alpha}).
       \end{split}
    \end{align}

It is worth noting that individual coherent components are not orthogonal, making the normalization factor of SCS depend on the coherent state amplitude $\alpha$. Moreover, orthogonality between $\ket{\mathcal{C}^0_\alpha}$ and $\ket{\mathcal{C}^1_\alpha}$ is strictly guaranteed by relative phase between their components regardless of their nonorthogonality. This can be explicitly proven by attaining the number basis expansion of 2-dimensional SCSs.
    \begin{align} \label{2d SCS}
        \begin{split}
           \bra{n}\ket{\mathcal{C}^0_\alpha} &= \frac{1}{\sqrt{2(1-e^{-2\alpha^2})n!}}(\alpha^n+(-\alpha)^n), \\
           \bra{n}\ket{\mathcal{C}^1_\alpha} &= \frac{1}{\sqrt{2(1-e^{-2\alpha^2})n!}}(\alpha^n-(-\alpha)^n).
        \end{split}
    \end{align}
Since $\ket{\mathcal{C}^0_\alpha}$ only contains number states with even photon number while $\ket{\mathcal{C}^1_\alpha}$ only contains the rest, they are orthogonal.

Generalized $d$-dimensional SCSs representing qudit basis can be obtained in the same way as the HES case.
    \begin{align}\label{SCS_def}
       \ket{\mathcal{C}^k_{\alpha,d}} := N_{k,\alpha,d} \sum_{n=0}^{d-1} \omega_d^{-kn} \ket{\alpha \omega_d^n},
    \end{align}
with qudit index $k$ and normalization factor $N_{k,\alpha,d}$ given by
    \begin{align} \label{SCS normalization factor}
       N_{k,\alpha,d} = \frac{1}{\sqrt{d \sum_{n=0}^{d-1} \omega_d^{-kn} \mathrm{exp}[-\alpha^2(1-\omega_d^n)]}}.
    \end{align}
    
Orthonormality is again guaranteed by the relative phase of each component canceling all the nonzero overlap of CV parts.
    \begin{align}    \label{SCS photon number}
    \begin{split}
      \bra{m} \ket{\mathcal{C}^k_{\alpha,d}} &= N_{k,\alpha,d} \sum_{n}\omega_d^{-kn} \bra{m}\ket{\alpha\omega_d^n} \\
      &= N_{k,\alpha,d} \sum_{n}\omega_d^{-kn}\frac{(\alpha\omega_d^n)^m}{\sqrt{m!}}
      \\&= \frac{\alpha^m N_{k,\alpha,d}}{\sqrt{m!}}\sum_{n}\omega_d^{n(m-k)} =  \frac{\alpha^m N_{k,\alpha,d}}{\sqrt{m!}}\delta_{m,k;d}.
    \end{split}
    \end{align}
    
Therefore, the qudit $\ket{\mathcal{C}^k_{\alpha,d}}$ consists only of states with $k(\mathrm{mod}\ d)$ photons. For this reason, SCS is also called pseudo-number state~\cite{Kim15}. This also reassures the orthogonality of SCS qudits.  

\section{Amplification of Hybrid Entangled states}
In this section, we compare the two amplification schemes applied to HESs. The first is addition-subtraction, namely $\hat{a}\hat{a}^\dagger$ scheme, while the second comprises successive photon addition, namely $a^\dagger{}^2$ scheme. Deterministic implementations of both schemes are fundamentally forbidden since they are non-unitary. 
Thus, quantum channels enabling the probabilistic application of those operations are designed to realize those schemes. 

The states after application of both schemes on $\ket{\mathcal{H}^k_{\alpha,d}}$ can be formally written as follows.
   \begin{align}
       \begin{split}
            &\ket{\mathcal{H}^{k,\hat{a}\hat{a}^\dagger}_{\alpha,d}} = N_{HES}^{\hat{a}\hat{a}^\dagger}(\alpha,d)\,(I \otimes \hat{a}{\hat{a}^\dagger})\frac{1}{\sqrt{d}}\sum_n \omega_d^{-kn} \ket{n, \alpha \omega_d^n}, \\
            &\ket{\mathcal{H}^{k,(\hat{a}^\dagger)^2}_{\alpha,d}} = N_{HES}^{(\hat{a}^\dagger{})^2} (\alpha,d)\,(I \otimes (\hat{a}^\dagger)^2 ) \frac{1}{\sqrt{d}}\sum_n \omega_d^{-kn} \ket{n, \alpha \omega_d^n},
       \end{split}
   \end{align}
where $N_{HES}^{\hat{a}\hat{a}^\dagger}(\alpha,d), N_{HES}^{(\hat{a}^\dagger)^2} (\alpha,d)$ normalization factors required for the consequent states to be normalized. Note that we only operate on CV parts.
    \begin{align} \label{hybrid 11 normalization}
       N_{HES}^{\hat{a}\hat{a}^\dagger}(\alpha,d) = \frac{1}{ \sqrt{\alpha^4 + 3\alpha^2 +1} },
    \end{align}
    \begin{align} \label{hybrid 02 normalization}
       N_{HES}^{(\hat{a}^\dagger)^2}(\alpha,d) = \frac{1}{ \sqrt{\alpha^4 + 4\alpha^2 +2} }.
    \end{align}
    
One can see that the normalization factors do not depend on the dimension $d$ nor qudit index $k$. This originates from the distinctive structure of HES where individual components are orthogonal to one another. This property of HES provides a way to simplify the calculation of relevant expectation values to a normal ordering problem of bosonic operators.

\noindent\textbf{Proposition 1} For any bosonic operator polynomial $P(\hat{a},\hat{a}^\dag)$ with equal degrees of $\hat{a}$ and $\hat{a}^\dag$ in every term, its expectation value to a HES $\ket{\mathcal{H}^k_{\alpha,d}}$ is same as the expectation value evaluated with a coherent state having same amplitude:
\begin{align} \label{general hybrid state}
    \langle P(\hat{a},\hat{a}^\dag) \rangle_{\mathcal{H}^k_{\alpha,d}} = \langle P(\hat{a},\hat{a}^\dag) \rangle_{\alpha}.
\end{align}

The trace preserving property of quantum channels and Proposition 1 give an expression for normalization factor $N_{HES}^{P(\hat{a},{\hat{a}^\dagger})}(\alpha,d)$:
\begin{align}
    N_{HES}^{P(\hat{a},{\hat{a}^\dagger})}(\alpha,d) = \frac{1}{\sqrt{\bra{\alpha}(P(\hat{a},{\hat{a}^\dagger}))^\dagger  P(\hat{a},{\hat{a}^\dagger})\ket{\alpha}}},
\end{align}
which is independent of dimension $d$ and qudit number $k$ as we observed. 

We take the fidelity and  the optimized gain as our figure of merit to compare the two schemes($\hat{a}\hat{a}^\dag, \hat{a}^\dagger{}^2$). The fidelity can be considered as a measure of how close we can get to our target state, while the gain quantifies the capability to increase the amplitude of the component states. Here, the fidelity is estimated between the initial state and the target state of the amplification. I.e., the target state is another HES with a larger coherent amplitude $\alpha$ and an additional phase factor from the amplification which results in a change of the qudit number $k$ as

\begin{equation}
    \begin{aligned}
        &F^{\hat{a}\hat{a}^\dag}_{HES} = \abs{\braket{\mathcal{H}^{k}_{g\alpha,d}}{\mathcal{H}^{k,\hat{a}\hat{a}^\dag}_{\alpha,d}}}^2, \\
        &F^{\hat{a}^\dagger{}^2}_{HES} = \abs{\braket{\mathcal{H}^{k+2}_{g\alpha,d}}{\mathcal{H}^{k,(\hat{a}^\dag)^2}_{\alpha,d}}}^2,
    \end{aligned}
\end{equation}
where $k+2$ for the target state of $F^{\hat{a}^\dagger{}^2}_{HES}$ is chosen to optimize the fidelity for the schme of two-photon addition.

The optimized gain, G, can be obtained by optimizing $g$ for the coherent amplitude achieving the highest fidelity. For the $\hat{a}\hat{a}^\dagger$ scheme, the fidelity is analytically obtained with respect to an arbitrary value of gain $g$ as
   \begin{align} \label{hybrid 11 fidelity}
        F^{\hat{a}\hat{a}^\dagger}_{HES} = \frac{g^2 \alpha^4 + 2 g \alpha^2 +1}{ \alpha^4 + 3\alpha^2 +1 } \mathrm{exp}[-\alpha^2(g-1)^2],
   \end{align}
while for the $(\hat{a}^\dag)^2$ scheme, the fidelity is obtained as
  \begin{align} \label{hybrid 02 fidelity}
        F^{\hat{a}^\dagger{}^2}_{HES} = \frac{g^4 \alpha^4 }{ \alpha^4 + 4\alpha^2 +2 } \mathrm{exp}[-\alpha^2(g-1)^2].
  \end{align}
  
\begin{figure}[t]
    \centering
    \begin{subfigure}[h]{\linewidth}
        
        \includegraphics[width=\textwidth]{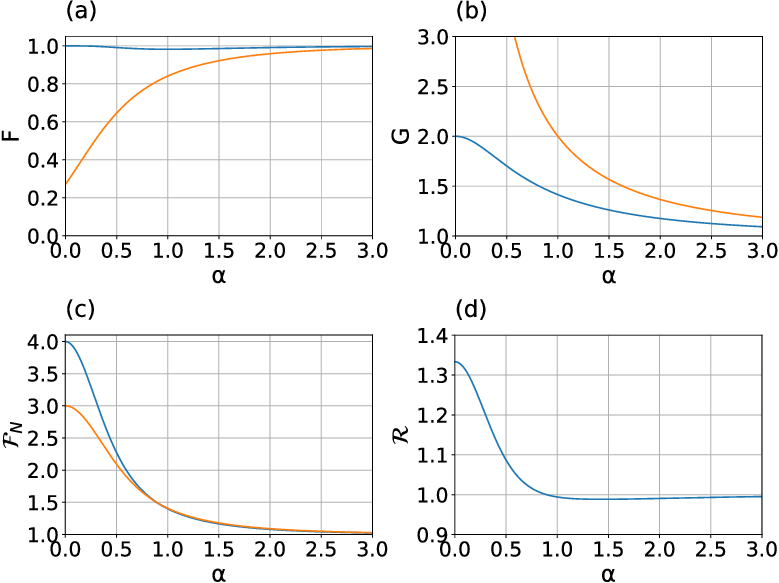}
    \end{subfigure}
    \caption{Several measures for the amplified HESs against input coherent amplitudes, $\alpha$, for comparison. (a) Fidelities F of the two schemes: $F_{HES}^{\hat{a}\hat{a}^\dagger}$(blue, solid) and $F_{HES}^{\hat{a}^\dagger{}^2}$(orange, dashed).
    (b) Optimized gains G of the two schemes: $G_{HES}^{\hat{a}\hat{a}^\dagger}$(solid blue) and $G_{HES}^{\hat{a}^\dagger{}^2}$(orange dashed).
    (c) Normalized quantum Fisher information, $\mathcal{F}_N$, of amplified HESs with respect to input HESs: $\mathcal{F}_{HES}^{\hat{a}\hat{a}^\dagger}/\mathcal{F}_{HES}$ (solid blue) and $\mathcal{F}_{HES}^{\hat{a}^\dagger{}^2}/\mathcal{F}_{HES}$(orange dashed).
    (d) Ratio of quantum Fisher information of the two schemes, $\mathcal{R}= \mathcal{F}_{HES}^{\hat{a}\hat{a}^\dagger} / \mathcal{F}_{HES}^{\hat{a}^\dagger{}^2}$.}
    \label{hybrid scheme comparison}
\end{figure}

The fidelities of the two schemes are plotted in Fig.~1(a). One can see that the $\hat{a}\hat{a}^\dag$ scheme yields higher fidelity than $(\hat{a}^\dag)^2$ scheme. We first note that the fidelities in Eqs.~(\ref{hybrid 11 fidelity}) and (\ref{hybrid 02 fidelity}) are the same with the fidelities between input and amplified coherent states due to Proposition 1. An explicit calculation of fidelities in Appendix A indeed has the form of Eq. (13). The behaviors of fidelities between the target coherent state with greater amplitude and the coherent state amplified with either the $\hat{a}\hat{a}^\dag$ scheme or $\hat{a}^\dagger{}^2$ scheme are  known~\cite{Park16}.

We can generalize this agreement between coherent states and HES qudits. Let us consider a general amplification scheme that employs successive photon addition and subtraction, which can be written as $\hat{a}^{n_1}\hat{a}^\dagger{}^{m_1}\hat{a}^{n_2}\hat{a}^\dagger{}^{m_2}\cdots$. We assume that $\sum_i (m_i - n_i) = l$, then it approximately maps $\ket{\mathcal{H}^k_{\alpha,d}}$ to $\ket{\mathcal{H}^{k+l}_{\beta,d}}$ for some $\beta$. The normal ordered form of this amplification would be the form of $\sum_i C_i \, \hat{a}^\dagger{}^{i} \hat{a}^\dagger{}^l \hat{a}^{i} = \hat{a}^\dagger {}^l :g(\hat{a},\hat{a}^\dagger):$ with every term in $g(\hat{a}, \hat{a}^\dag)$ containing the same number of creation/annihilation operators, thus an overlap between the amplified state and the target state is written as

\begin{align}
\begin{split}
&\langle \mathcal{H}_{\beta, d}^{k+l}|\hat{a}^\dagger {}^l :g(\hat{a},\hat{a}^\dagger):|\mathcal{H}_{\alpha, d}^{k}\rangle \\
    \\
    = &\frac{1}{d}\sum_{m,n} \bra{m, \beta \omega^m} (I \otimes \hat{a}^\dagger{}^l :g(\hat{a},\hat{a}^\dagger):\,) \ket{n, \alpha \omega^n} \omega^{l(m-n)} \\
    = &\frac{1}{d} \sum_n \,\beta^l \,\bra{\beta \omega^n} :g(\hat{a},\hat{a}^\dagger):\ket{\alpha \omega^n}     =\beta^l \bra{\beta}:g(\hat{a},\hat{a}^\dagger):\ket{\alpha},
\end{split}
\end{align}
which is exactly that of coherent states. That is, we can generalize Proposition 1 further:

\noindent\textbf{Proposition 2} For any bosonic operator polynomial $Q(\hat{a}, \hat{a}^\dag;l)$ with every term in it satisfying $\mathrm{deg}(\hat{a})-\mathrm{deg}(\hat{a}^\dag)=l$,
\begin{equation}
    \langle \mathcal{H}_{\beta, d}^{k+l}|Q(\hat{a}, \hat{a}^\dag;l)|\mathcal{H}_{\alpha, d}^{k}\rangle = \langle\beta|Q(\hat{a}, \hat{a}^\dag;l)|\alpha\rangle.
\end{equation}

Unlike Proposition 1, this agreement between the fidelity of the HES amplification and that of the coherent-state amplification does not solely originate from the orthogonality of discrete variables. We both need discrete variables $\ket{n}$ which act as labels of continuous variables $\ket{\alpha \omega_d^n}$ and equal phase factor $\omega$ which constitutes both qudit number $k$ by relative phase $\omega^{-kn}$ and phase of coherent states $\ket{\alpha \omega^n}$ to cancel out relative phase.

We now obtain the optimized gain defined as the ratio between the amplitudes of the target and initial states giving the maximum fidelity. This can be obtained by solving the equation
   \begin{align} \label{def of gain}
      \dv{F}{g} \bigg |_{g=\rm{G}}=0.
   \end{align}
For the $\hat{a}\hat{a}^\dag$ scheme, solving the equation gives the optimized gain as
   \begin{align}
       G_{HES}^{\hat{a}\hat{a}^\dag} = \frac{\alpha^2 -1 + \sqrt{\alpha^4 + 6 \alpha^2 +1}}{2 \alpha^2},
   \end{align}
while for the $(\hat{a}^\dag)^2$ scheme, the optimized gain is obtained as
   \begin{align}
       G_{HES}^{\hat{a}^\dagger{}^2} = \frac{1 + \sqrt{1+\frac{8}{\alpha^2}}}{2}.
   \end{align}
The optimized gains of the two schemes are shown in Fig. 1(b). One can see that the $\hat{a}^\dagger{}^2$ scheme yields the higher optical gain than the $\hat{a}\hat{a}^\dag$ scheme in contrast with the fidelities. The behavior of optimized gain of hybrid state amplification is also identical to that of coherent state amplification for the same reason as the case of the fidelity.

We further note that the gain can be defined by the ratio of the expectation values of quadrature operators with an arbitrary phase $\phi$~\cite{Zavatta11, Park16}. This gain value, however, can be somewhat inadequate for HESs because the expectation value of quadrature value with any phase $\phi$ is zero for HESs, as one can confirm with the equation
\begin{align} \nonumber
    &\frac{1}{d}\sum_{m,n} \bra{m, \alpha\omega^m} I \otimes (\hat{a}e^{i\lambda}+\hat{a}^\dagger e^{-i\lambda}) \ket{n, \alpha \omega^n} \\
    =&\frac{1}{d} \sum_n \alpha (e^{i\lambda} \omega^n + e^{- i\lambda} \omega^{-n}) = 0.
\end{align}
Interestingly, however, the ratio is finite and equals that of coherent states. For an arbitrary amplification scheme via photon addition or subtraction, $f(\hat{a},{\hat{a}^\dagger})$, the gain of coherent states defined by the ratio is
\begin{align} \nonumber
    g = (N^{f(\hat{a},{\hat{a}^\dagger})})^2\frac{ \bra{\alpha} f(\hat{a},{\hat{a}^\dagger})^\dagger (\hat{a} e^{i\lambda} + \hat{a}^\dagger e^{-i\lambda}) f(\hat{a},{\hat{a}^\dagger}) \ket{\alpha}}
    {\bra{\alpha} (\hat{a} e^{i\lambda} + \hat{a}^\dagger e^{ -i\lambda}) \ket{\alpha}}.
\end{align}
If we write the normal ordered form of $f(\hat{a},{\hat{a}^\dagger})^\dagger \hat{a} f(\hat{a},{\hat{a}^\dagger})$ as $\sum_i C_i \, \hat{a}^\dagger{}^{i} \hat{a}^\dagger{}^i \hat{a} = g(\hat{a},\hat{a}^\dagger)\hat{a}$, the gain is expressed as
\begin{align} \nonumber
    g = (N^{f(\hat{a},{\hat{a}^\dagger})})^2 \bra{\alpha} g(\hat{a},\hat{a}^\dagger) \ket{\alpha}.
\end{align}
Likewise, the gain of HESs defined by the ratio is
\begin{align} \nonumber
    g_{HES} = (N_{HES}^{f(\hat{a},{\hat{a}^\dagger})})^2 \bra{\alpha} g(\hat{a},\hat{a}^\dagger) \ket{\alpha}.
\end{align}
We can go further with this gain to define the  EIN~\cite{Grangier92}, a measure of noiseless amplification. However, since the expectation value of quadrature is zero, it lacks physical meaning, and we will not stick to it.

Observing the results so far, the $\hat{a}\hat{a}^\dag$ scheme has an advantage of higher fidelity while $\hat{a}^\dagger{}^2$ has an advantage of higher gain, for any amplitude $\alpha$. This property provides us with a guide of which amplification scheme should be used. However, their advantages in different aspects make it hard to compare their overall performance. Qualitatively, we know that the $\hat{a}\hat{a}^\dag$ scheme would be better for small amplitudes since coherent states with small amplitudes are susceptible to the change of their photon numbers. Meanwhile, we can guess that the two schemes will not differ so much for large amplitudes because coherent states with sufficiently large amplitudes are robust against several photon addition or subtraction processes. Lastly, as a consequence, we can deduce that the $\hat{a}^\dagger{}^2$ scheme would be appropriate for the intermediate amplitude region.

A previous work~\cite{Park16} employed two measures, the EIN for coherent states and the quantum Fisher information for $2$-dimensional SCSs to quantify overall performance which reflects aforementioned qualitative arguments. The EIN is a measure of undesired noise generated in the amplification process, thus it is appropriate to measure overall performance. However, we have concluded that the EIN lacks physical meaning in our case since the expectation values of quadrature operators are always zero. 

As an alternative, we take the quantum Fisher information 
associated with the Hamiltonian of the harmonic oscillator $\hat{H}$. The quantum Fisher information is then defined as
\begin{align}
    \mathcal{F}[\hat{\rho},\hat{H}] = 2 \sum_{k,l}\frac{(\lambda_k-\lambda_l)^2}{(\lambda_k+\lambda_l)}|\bra{k}\hat{H}\ket{l}|^2,
\end{align}
where $\hat{H}:=\hat{a}^\dagger\hat{a}$, $\hat{\rho}$ is a density matrix of a state used for the estimation of parameter $\theta$,  $\ket{k}$ is a $k$-th eigenvector, and $\lambda_k$ is a $k$-th eigenvalue of the density matrix $\rho$.
Here, we use a parametrized density matrix $\hat{\rho}(\theta):=e^{-i\hat{H}\theta}\hat{\rho}e^{i\hat{H}\theta}$ for the phase estimation. We only consider the bosonic mode even for the HES to make a fair comparison. 
The quantum Fisher information is in a direct relationship with the performance of phase estimation. The quantum Cram\'er-Rao bound, 
\begin{equation}
    \mathrm{var}(\theta) \geq \frac{1}{\mathcal{F}(\theta)},
\end{equation}
imposes a lower bound on the statistical error of the phase estimation scheme.

If the state of interest is pure, the quantum Fisher information is easily given by $4 \langle \Delta \hat{H} \rangle $, which is proportional to $4 \langle \Delta \hat{n} \rangle$ for optical systems, where $\hat{n}$ is the number operator $\hat{a}^\dagger \hat{a}$. For simplicity, we will call $4 \langle \Delta \hat{n} \rangle$ as the quantum Fisher information from now on. To begin with, the quantum Fisher information of the HES is
    \begin{align}
            &\mathcal{F}_{HES} = 4\alpha^2,
    \end{align}
which is again the same as that of the coherent state due to Proposition 1. In the same argument of Ref.~\cite{Park16}, this supports that the quantum Fisher information is a good measure of the overall performance of the amplification schemes because both amplitude gain and noiseless property imply larger quantum Fisher information. The quantum Fisher information of $\hat{a}\hat{a}^\dag$-amplified hybrid states is also given by
    \begin{align}
            \mathcal{F}_{HES}^{\hat{a}\hat{a}^\dagger} = \frac{4\alpha^2(\alpha^8+6\alpha^6+14\alpha^4+10\alpha^2+4)}{(\alpha^4+3\alpha^2+1)^2},
    \end{align}
and the quantum Fisher information of $\hat{a}^\dagger{}^2$-amplified hybrid states is 
    \begin{align}
           \mathcal{F}_{HES}^{\hat{a}^\dagger{}^2} = \frac{4\alpha^2(\alpha^8+8\alpha^6+24\alpha^4+24\alpha^2+12)}{(\alpha^4+4\alpha^2+2)^2}.
    \end{align}

The ratios of quantum Fisher information, $\mathcal{F}_{HES}^{\hat{a}\hat{a}^\dagger}/\mathcal{F}_{HES}$ and $\mathcal{F}_{HES}^{\hat{a}^\dagger{}^2}/\mathcal{F}_{HES}$, are plotted in Fig. 1(c). The value is always larger than $1$, indicating that the HESs never become worse in terms of the quantum Fisher information by the amplification schemes. In the asymptotic limit, the ratio converges to $1$, implying that the amplification schemes are trivial in such a regime. 

For a clear comparison of both schemes, we examined the ratio between two schemes $\mathcal{F}_{HES}^{\hat{a}\hat{a}^\dagger}/\mathcal{F}_{HES}^{\hat{a}^\dagger{}^2}$, in Fig. 1(d). The figure shows that up to amplitude $\alpha \approx 0.9$, the $\hat{a}\hat{a}^\dag$ scheme has the higher quantum Fisher information. The ratio monotonically decreases to $\alpha \approx 1.43$, starting from $1.33$ at $\alpha = 0$, $1.11$ at $\alpha = 0.45$, and $1 $ at $\alpha \approx 0.9$. After $\alpha \approx 0.9$, the $\hat{a}^\dagger{}^2$ scheme yields the higher quantum Fisher information for the amplified state. For sufficiently large $\alpha$, the ratio converges to $1$ and the difference is negligible. This result confirms our qualitative analysis.

\section{Amplification of Superpositions of Coherent States}

In this section, we compare the two amplification schemes performed on SCSs. After the $\hat{a}\hat{a}^\dag$ scheme and $\hat{a}^\dagger{}^2$ scheme, respectively,
the amplified SCSs can be expressed as
\begin{equation}   
   \begin{aligned}
       &\ket{\mathcal{C}_{\alpha, d}^{k, \hat{a}\hat{a}^\dagger}} := N_{SCS}^{\hat{a}\hat{a}^\dagger}(\alpha,k,d)\, \hat{a}{\hat{a}^\dagger}\sum_n \omega_d^{-kn} \ket{\alpha \omega_d^n}, \\
       &\ket{\mathcal{C}_{\alpha, d}^{k, (\hat{a}^\dagger)^2}} :=
        N_{SCS}^{(\hat{a}^\dagger{})^2}(\alpha,k,d)\,(\hat{a}^\dagger)^2  \sum_n \omega_d^{-kn} \ket{\alpha \omega_d^n},
   \end{aligned}
\end{equation}  
where $N_{SCS}^{\hat{a}\hat{a}^\dagger}(\alpha,k,d)$ and $N_{SCS}^{(\hat{a}^\dagger{})^2}(\alpha,k,d)$ are normalization factors
   \begin{widetext}
      \begin{align}
        &N_{SCS}^{\hat{a}\hat{a}^\dagger}(\alpha,k,d) = \frac{1}{\sqrt{d \sum_n \omega_d^{-kn} (\alpha^4 \omega_d^{2n} + 3\alpha^2 \omega_d^n +1) \mathrm{exp}[-\alpha^2(1-\omega_d^n)]}},\\
        &N_{SCS}^{(\hat{a}^\dagger{})^2}(\alpha,k,d) = \frac{1}{\sqrt{d \sum_ n \omega_d^{-kn} (\alpha^4 \omega_d^{2n} + 4\alpha^2 \omega_d^n +2) \mathrm{exp}[-\alpha^2(1-\omega_d^n)]}}.
      \end{align}
   \end{widetext}
Both the normalization factors $N_{SCS}^{(\hat{a}^\dagger{})^2}(\alpha,k,d)$ and $N_{SCS}^{(\hat{a}^\dagger{})^2}(\alpha,k,d)$  depend on the qudit number $k$ and dimension $d$, unlike the normalization factors of the amplified HESs. Explicit forms of the amplified SCSs are presented in Appendix B.

As  explained in Sec.~III with Eq.~(15), we first obtain the fidelities and the optimized gains defined by the values that maximize the fidelities:
\begin{equation}
    \begin{aligned}
        &F^{\hat{a}\hat{a}^\dagger}_{SCS} := \abs{\braket{\mathcal{C}^{k}_{g\alpha, d}}{\mathcal{C}^{k, \hat{a}\hat{a}^\dagger}_{\alpha, d}}}^2, \\
        &F^{\hat{a}^\dagger{}^2}_{SCS} := \abs{\braket{\mathcal{C}^{k+2}_{g\alpha, d}}{\mathcal{C}^{k, (\hat{a}^\dagger)^2}_{\alpha, d}}}^2.
    \end{aligned}
\end{equation}
For each scheme, the fidelity can be expressed as a function of $\alpha$ and $g$ as 
   \begin{widetext}
        \begin{align}
             F_{SCS}^{\hat{a}\hat{a}^\dagger} = \frac{\{\sum_n (1+g\alpha^2\omega_d^{n}) \omega_d^{-kn} \mathrm{exp}[g\alpha^2 \omega_d^n] \}^2}{(\sum_n \omega_d^{-kn}(\alpha^4 \omega_d^{2n} + 3\alpha^2\omega_d^n+1)\mathrm{exp}[\alpha^2\omega_d^n])(\sum_n \omega_d^{-kn}\mathrm{exp}[g^2\alpha^2\omega_d^n]) },
         \end{align}

        \begin{align}
             F_{SCS}^{\hat{a}^\dagger{}^2} = \frac{g^4\alpha^4 \{ \sum_n \omega_d^{- k n} \mathrm{exp}[g\alpha^2 \omega_d^n] \}^2}{(\sum_n \omega_d^{-kn}(\alpha^4 \omega_d^{2n} + 4\alpha^2\omega_d^n +2)\mathrm{exp}[\alpha^2\omega_d^n])(\sum_n \omega_d^{-(k+2)n}\mathrm{exp}[g^2\alpha^2\omega_d^n]) }.
         \end{align}
   \end{widetext}
The optimized gains are then obtained to maximize the fidelities:
   \begin{align}
       \dv{{F_{SCS}^{\hat{a}\hat{a}{}^\dagger}}}{g}\bigg |_{g=G^{\hat{a}\hat{a}^\dagger}_{SCS}}=0, \,\,\,\, \dv{{F_{SCS}^{\hat{a}^\dagger{}^2}}}{g}\bigg |_{g=G^{(\hat{a}^\dagger)^2}_{SCS}}=0.
   \end{align}
We find numerical values of the gain using the downhill simplex algorithm since analytic forms cannot be obtained in closed forms using elementary functions. The implementation was performed through the Python package Scipy, using the ~\textbf{scipy fmin} function. Relations between the fidelity and gain for both the schemes are shown in Figs. 2 and 3 for $d=3, 4, 5$ with every $k$. In Fig. 2, the fidelity and the optimized gain of the $\hat{a}\hat{a}^\dagger$ amplification on SCSs are plotted against the coherent amplitude $\alpha$. Behaviors of those quantities do not differ much from the HES case, however, there are two additional features. (i) The larger the qudit number $k$, the larger the amplitude $\alpha$ where the local minimum appears. (ii) The larger the qudit number $k$, the larger the local minimum.

\begin{figure}[h]
    \centering
    \begin{subfigure}[h]{\linewidth}
        \centering
        \includegraphics[width=\textwidth]{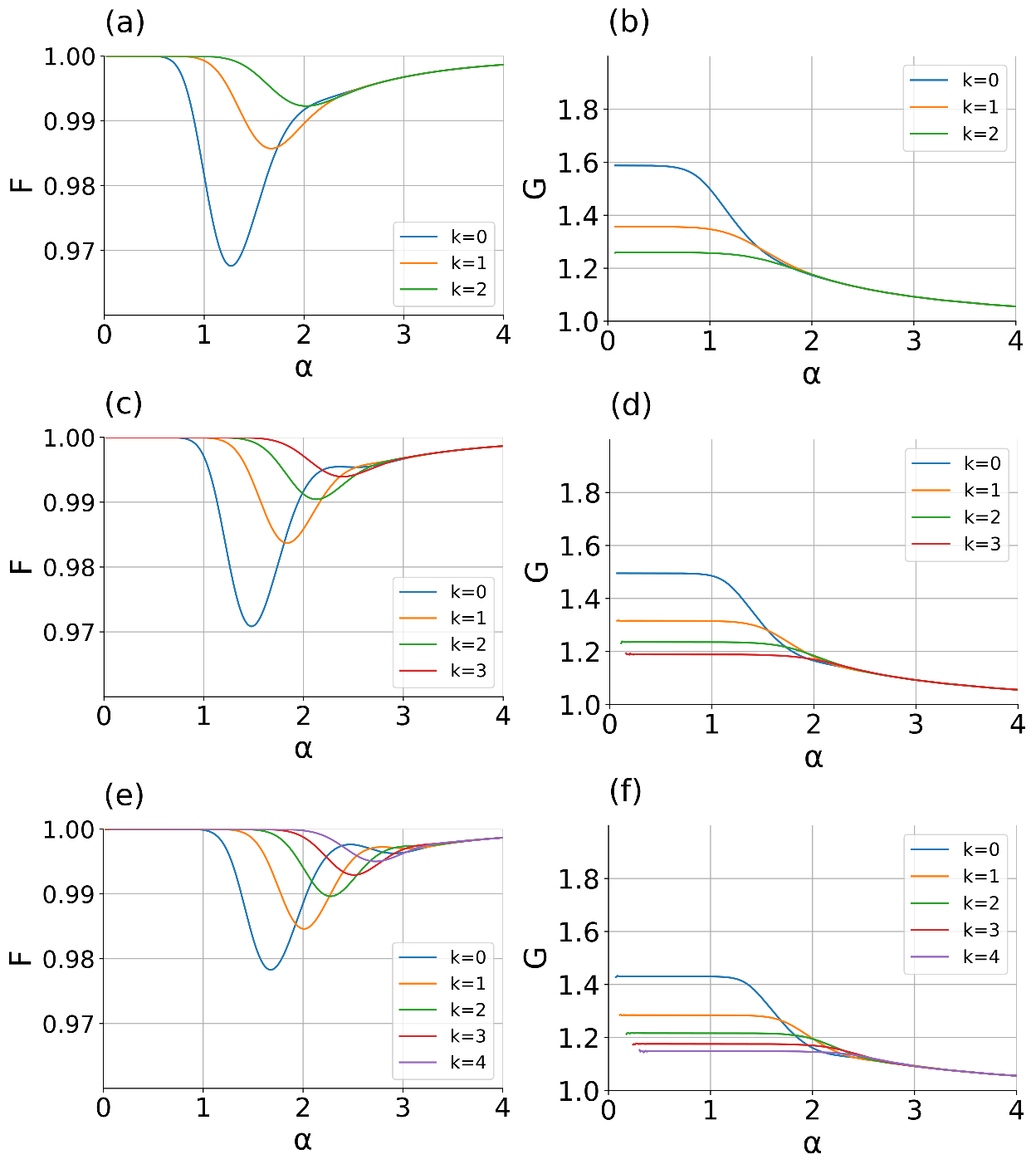}
    \end{subfigure}
    \caption{Fidelity (F) and optimized gain (G) obtained from the $\hat{a}\hat{a}^\dagger$ scheme against input coherent amplitude $\alpha$ of SCSs with qudit dimension [(a), (b) $d=3$, (c),(d) $d=4$, (e), (f) $d=5$], respectively. Here, $k$ is the integer qudit index in Eq.~\ref{SCS_def} with the relative phase information of the $d$-dimensional qudit.}
    \label{SCS fidelity gain comparison}
\end{figure}

\begin{figure}[h]
    \centering
    \begin{subfigure}[h]{\linewidth}
        \includegraphics[width=\textwidth]{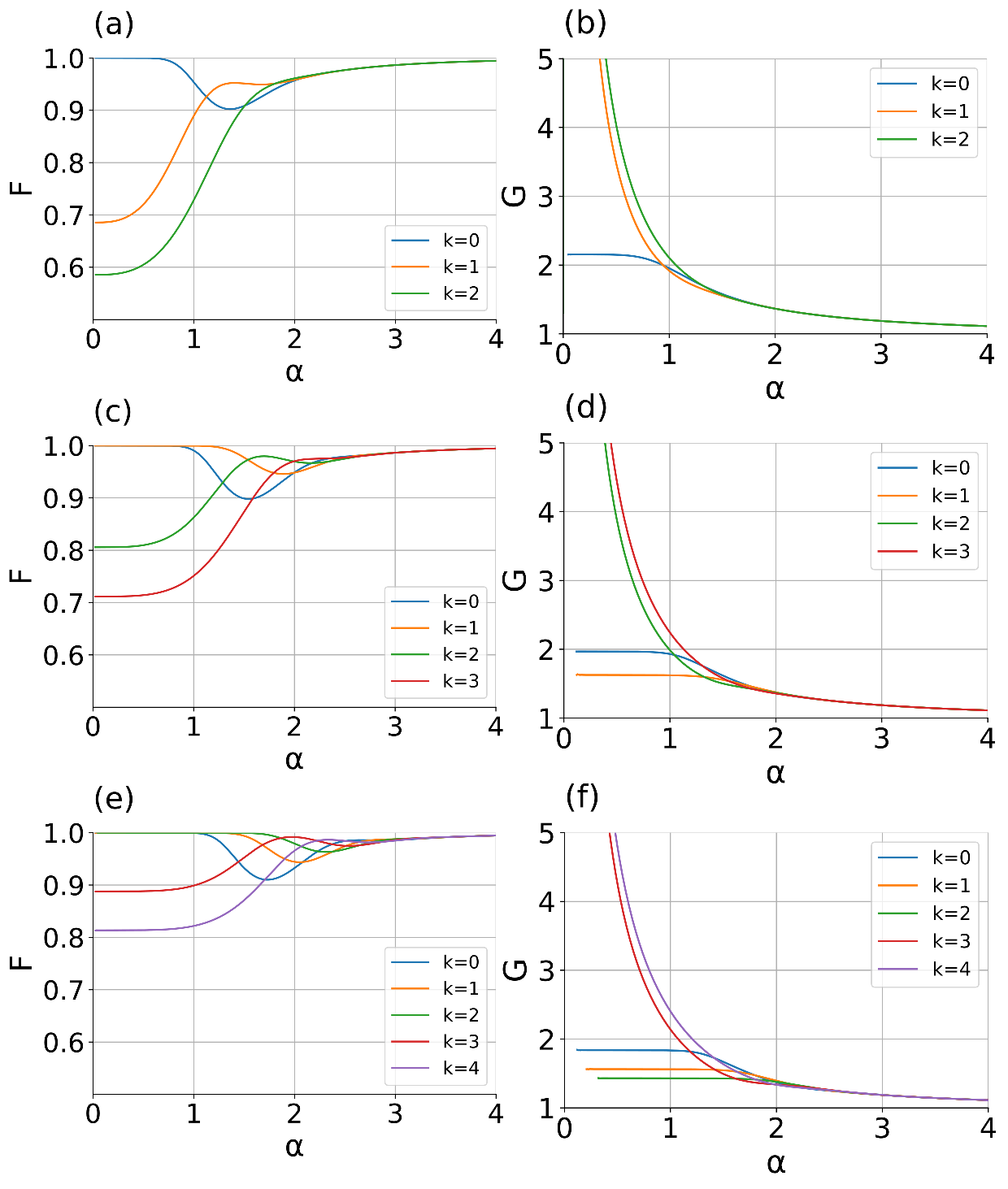}    
    \end{subfigure}
    \caption{Fidelity (F) and optimized gain (G) obtained from the $\hat{a}^\dagger{}^2$ scheme against coherent amplitude $\alpha$ of input SCSs with qudit dimension [(a), (b) $d=3$, (c), (d) $d=4$, (e), (f) $d=5$], respectively. }
    \label{SCS fidelity gain comparison}
\end{figure}

We recall that a SCS with qudit number $k$, $\ket{\mathcal{C}^k_{\alpha,d}}$, contains $k~(\mathrm{mod}\ d)$ number of photons. For a small amplitude, the coefficient of Fock basis $\ket{n}$ is concentrated on the Fock state with the least number, $\ket{k}$. This explains high fidelity for very low amplitude because the amplification $\hat{a}\hat{a}^\dagger$ does not change the number state, as one can confirm with the following equality
\begin{align}
    \hat{a}\hat{a}^\dagger \ket{k} = (k+1) \ket{k}.
\end{align}

As $\alpha$ increases, the SCS is no longer an approximation of a number state of $k$. State $\ket{\mathcal{C}^k_{\alpha,d}}$ contains more Fock states such as $\ket{d+k}$, $\ket{2d+k}$, .... The additional factor arising from the amplification then becomes relevant and changes the state as
\begin{align} \nonumber
    \hat{a}\hat{a}^\dagger \,&(\,\ket{k} + \ket{d+k}\,)  = (k+1)\ket{k}+(d+k+1)\ket{d+k}\\
    \not\propto &(\,\ket{k} + \ket{d+k}\,).
\end{align}
This explains the emergence of troughs as $\alpha$ increases. If amplitude $\alpha$ is sufficiently large, every $\ket{\alpha \omega_d^n}$ is orthogonal to each other as one can see from the equation
\begin{align}
    \bra{\alpha \omega_d^m} \ket{\alpha \omega_d^n} = \mathrm{exp}[-\alpha^2(1-\omega_d^{(n-m)})] \longrightarrow \delta_{m,n},
\end{align}
as $\alpha \longrightarrow \infty$. Accordingly, the normalization factor $N_{k,\alpha,d}$ converges to $\frac{1}{\sqrt{d}}$ as $\alpha$ grows, thus the difference between the SCSs and hybrid states vanishes. This explains the recovery of high fidelity for large amplitude.

We revisit the pseudo-number property of SCSs to further investigate the behavior of their fidelities against amplified counterparts. 
It is clear that the mean photon numbers of the SCSs become larger as the amplitudes $\alpha$ increase as also seen in Eq.~(\ref{SCS photon number}). While the physical interpretation of the amplitude $\alpha$ renders this intuitive, one can verify it by examining the sign of the derivative. As we mentioned earlier, $\ket{\mathcal{C}^k_{\alpha,d}}$ starts from the Fock state $\ket{k}$ for $\alpha \sim 0$, and obtain coefficients of other Fock states $\ket{d+k}$, $\ket{2d+k}$, $\dots$, as $\alpha$ increases. For small qudit-number $k$, coefficients of other Fock states can be sufficiently large to reduce the fidelity with small amplitude $\alpha$. This explains why trough appears earlier with smaller qudit-number $k$. As amplitude $\alpha$ grows, the overlap between $\{\ket{\alpha \omega_d^n}\}_n$ becomes smaller, and the recovered orthogonality contributes to the fidelity which makes the depth of trough shallower as $k$ increases.

With this qualitative analysis, we can predict two additional features that were not seen in Fig. 2. First, since the overlap between $\{\ket{\alpha \omega_d^n}\}_n$ increases with $d$, we need larger amplitude $\alpha$ to recover the orthogonality. The coefficients of $\ket{2d+k}, \ket{3d+k}, \dots$, can then be significantly large before we recover the orthogonality, and it makes multiple troughs as shown in the fidelity plots of Fig.~3. Second, for fixed $k$, there is no monotonic tendency of the depth of troughs concerning $d$ because two factors antagonize each other. If $d$ grows, the overlap between $\{\ket{\alpha \omega_d^n}\}_n$ increases, but the required amplitude $\alpha$ to see trough becomes large, contributing to orthogonality. One can see that the depth of troughs of qudit number $0$ with $d=3$ is deeper than that of $d=2$ and $d=4$.

The gain graphs of Fig.~3 do not show any unusual behaviors. The SCSs have more photons as $\alpha$ increases, and the contribution to the gain for the $\hat{a}\hat{a}^\dag$ scheme monotonicall decreases with $\alpha$.

In Fig.~3, we plot the fidelity and the optimized gain against $\alpha$ for the $\hat{a}^\dagger{}^2$ scheme. The most noticeable feature is that only two states show unusual behaviors for the fidelity and the gain. We focus on the fact that the qudit-numbers of these two states are the last two, that is, $d-2$ and $d-1$. Recall that a qudit number $k$ is mapped to $k+2$ (mod $d$) via $\hat{a}^\dagger{}^2$ amplification, thus our target states are $\ket{\mathcal{C}^0_{\beta,d}}$ and $\ket{\mathcal{C}^1_{\beta,d}}$. For sufficiently large $\alpha$, it does not show any unusual behavior due to the orthogonality. However, for small amplitudes, $\alpha \sim 0$, since $\ket{\mathcal{C}^{(d-2)}_{\alpha,d}} \sim \ket{d-2}$ and $\ket{\mathcal{C}^{(d-1)}_{\alpha,d}} \sim \ket{d-1}$, the $\hat{a}^\dagger{}^2$ scheme maps most of their coefficients to $\ket{d}$ and $\ket{d+1}$, which do not have a large overlap with $\ket{\mathcal{C}^0_{\alpha,d}} \sim \ket{0}$ and $\ket{\mathcal{C}^1_{\alpha,d}} \sim \ket{1}$. This makes low fidelity for the states with qudit-numbers $d-2$ and $d-1$. Moreover, the target states should have large amplitude $\alpha$ to have coefficients of $\ket{d}$ and $\ket{d+1}$, thus the ratio between the amplitude of the target state and that of the initial state diverges, which explains the behavior of the gain graphs for small $\alpha$. As $\alpha$ increases, target states have larger coefficients of $\ket{d}, \ket{2d}, \dots$ and $\ket{d+1}, \ket{2d+1}, \cdots$, thus their fidelity increase and gain values decrease. Furthermore, for small amplitude $\alpha$, with the same reason of troughs appears in $\hat{a}\hat{a}^\dagger$ amplification, fidelity is higher for $\ket{\mathcal{C}^{(d-2)}_{\alpha,d}}$, and thus the gain is higher for $\ket{\mathcal{C}^{(d-1)}_{\alpha,d}}$. As $\alpha$ increases, they can be reversed as one can see in Fig. 3, (e) and (f) with $\alpha \approx 2.2$. Other states except for the last two show similar behavior to Fig. 2. Their troughs are deeper than those of $\hat{a}\hat{a}^\dagger$ state because $\hat{a}^\dagger{}^2$ changes photon number more than $\hat{a}\hat{a}^\dagger$. This makes lower fidelity and higher gain, which is also observed for HESs or coherent states~\cite{Park16}.

Now, we investigate the quantum Fisher information of the SCSs and the amplified SCSs to compare their performance of optimal phase estimation, as we did in section III. The quantum Fisher information of SCSs and amplified SCSs under $\hat{a}\hat{a}^\dagger$ and $\hat{a}^\dagger{}^2$ schemes is calculated to be
   \begin{widetext}
        \begin{align}
            \mathcal{F}_{SCS} = 4\alpha^2 \frac{ \sum_n (\alpha^2 \omega_d^{(2-k)n} + \omega_d^{(1-k)n}) \mathrm{exp}[\alpha^2\omega_d^n]}{ \sum_n \omega_d^{-kn} \mathrm{exp}[\alpha^2\omega_d^n]} - 4\alpha^4 \left(\frac{\sum_n \omega_d^{(1-k)n} \mathrm{exp}[\alpha^2\omega_d^n]}{ \sum_n \omega_d^{-kn} \mathrm{exp}[\alpha^2\omega_d^n]}\right)^2,
        \end{align}

       \begin{align}\nonumber
            \mathcal{F}_{SCS}^{\hat{a}\hat{a}^\dagger} =& \frac{ 4\sum_n (\alpha^8 \omega_d^{4n} +8 \alpha^6 \omega_d^{3n} +14 \alpha^4 \omega_d^{2n} + 4 \alpha^2 \omega_d^{n}) \mathrm{exp}[\alpha^2\omega_d^n] \omega_d^{-kn}}{ \sum_n (\alpha^4 \omega_d^{2n} + 3 \alpha^2 \omega_d^{n} + 1 )\omega_d^{-kn} \mathrm{exp}[\alpha^2\omega_d^n]} \\
            &- 4\left( \frac{ \sum_n (\alpha^6 \omega_d^{3n} +5 \alpha^4 \omega_d^{2n} + 4 \alpha^2 \omega_d^{n}) \mathrm{exp}[\alpha^2\omega_d^n] \omega_d^{-kn}}{ \sum_n (\alpha^4 \omega_d^{2n} + 3 \alpha^2 \omega_d^{n} +1 )\omega_d^{-kn} \mathrm{exp}[\alpha^2\omega_d^n]} \right)^2,
       \end{align}
       \begin{align} \nonumber
           \mathcal{F}_{SCS}^{\hat{a}^\dagger{}^2} =& \frac{ 4\sum_n (\alpha^8 \omega_d^{4n} +13 \alpha^6 \omega_d^{3n} +46 \alpha^4 \omega_d^{2n} + 46 \alpha^2 \omega_d^{n} +8) \mathrm{exp}[\alpha^2\omega_d^n] \omega_d^{-kn}}{ \sum_n (\alpha^4 \omega_d^{2n} + 4 \alpha^2 \omega_d^{n} +2 )\omega_d^{-kn} \mathrm{exp}[\alpha^2\omega_d^n]} \\
            &- 4\left( \frac{ \sum_n (\alpha^6 \omega_d^{3n} +8 \alpha^4 \omega_d^{2n} + 14 \alpha^2 \omega_d^{n} +4) \mathrm{exp}[\alpha^2\omega_d^n] \omega_d^{-kn}}{ \sum_n (\alpha^4 \omega_d^{2n} + 4 \alpha^2 \omega_d^{n} +2 )\omega_d^{-kn} \mathrm{exp}[\alpha^2\omega_d^n]} \right)^2.
        \end{align}
    \end{widetext}

One can see that the quantum Fisher information of $d$-dimensional SCSs is not a monotonic function with respect to amplitude $\alpha$, thus we cannot follow the argument in  Sec. III and  Ref.~\cite{Park16} that Fisher information reflects amplification gain and noiseless property. In Fig. 4, we plotted quantum Fisher information of SCSs for $d=2, 3, 4$ and $5$. For $d=2$, the graphs smoothly increase with amplitude, and for $d=3$, the graph shows some fluctuation but still three graphs monotonically increase with amplitude. However, for $d=4$ and $d=5$, the fluctuation width becomes so large that they lose monotonic behavior. This shows that the quantum Fisher information is not a good measure of the overall performance of noiseless amplification for high dimensions. Smooth and monotonic behavior of quantum Fisher information of SCSs only appears at $d=2$, even and odd cat states, which were the main interest of  Ref.~\cite{Park16}. Therefore, the quantum Fisher information for $d$-dimensional SCSs should be considered as a performance of optimal phase estimation, one of the criteria to compare with other optical states since phase estimation is a popular task in quantum optics, while it is a good measure of overall performance of amplification for hybrid states and $d=2$ SCSs.

\begin{figure}[b]
    \centering
    \begin{subfigure}[h]{\linewidth}
        \centering
        \includegraphics[width=\textwidth]{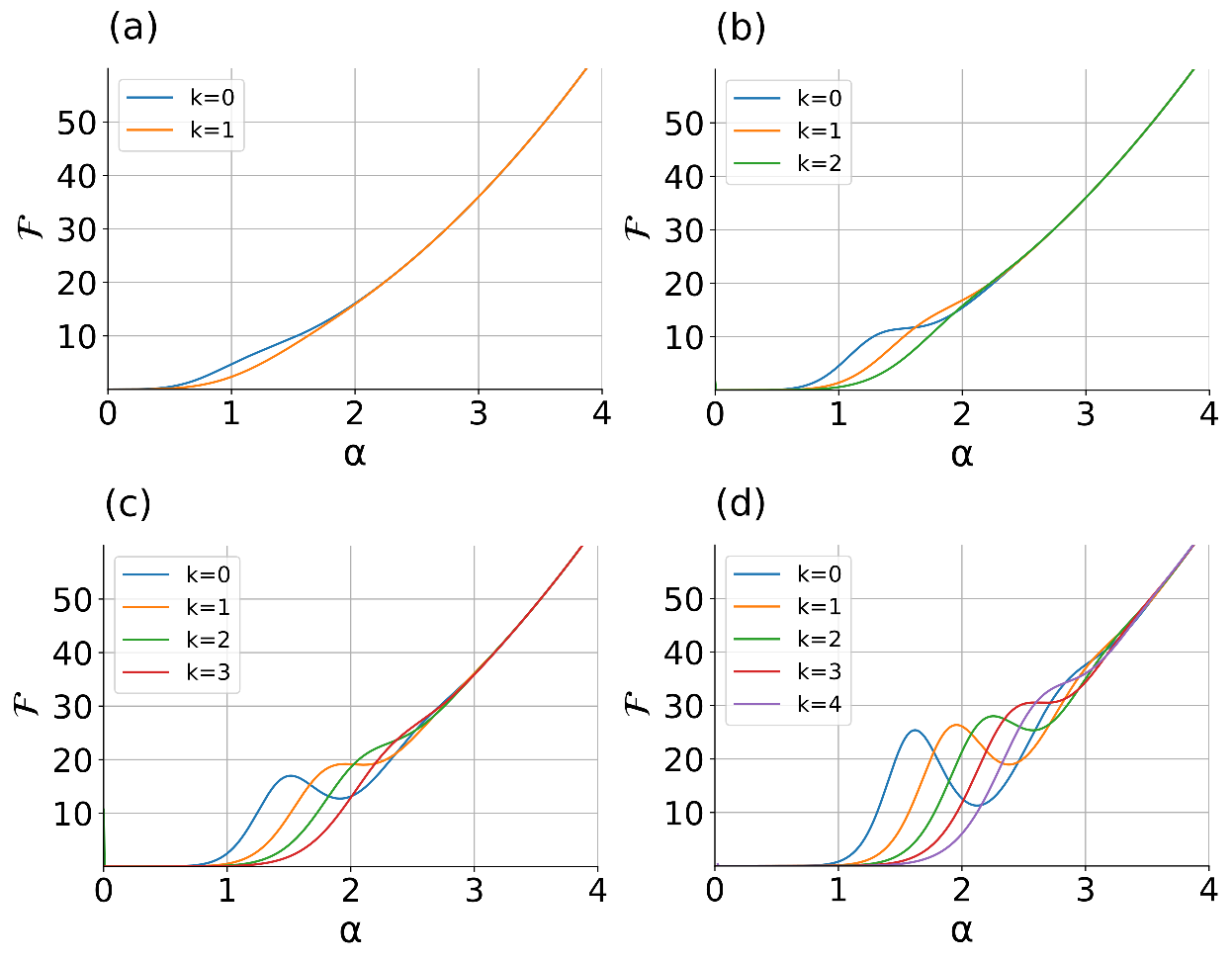}
    \end{subfigure}
    \caption{Quantum Fisher information of SCSs($\mathcal{F}_{SCS}$) with coherent amplitudes $\alpha$ for (a) $d=2$, (b) $d=3$, (c) $d=4$, and (d) $d=5$.}
    \label{SCS Fisher Information}
\end{figure}

We analyze the fluctuating behavior of quantum Fisher information of SCSs with pseudo-number property. First, observe that Fock states are invariant under phase rotation. This implies that we cannot use the Fock state for phase estimation, which is reflected in quantum Fisher information that $\mathcal{F}_{\ket{n}} = 0$ for any Fock state $\ket{n}$. As we investigated, $\ket{\mathcal{C}^k_{\alpha,d}} \sim \ket{k}$ for $\alpha \sim 0$, thus their quantum Fisher information are nearly zero. As amplitude grows, $\ket{\mathcal{C}^k_{\alpha,d}}$ is no longer rotationally invariant thus its quantum Fisher information increases. As amplitude grows further where $\ket{\mathcal{C}^k_{\alpha,d}}$ begins to concentrate to some Fock state $\ket{d\cdot n+k}$, its quantum Fisher information begins to decrease. Further growing amplitude makes $\ket{\mathcal{C}^k_{\alpha,d}}$ be a superposition of Fock states again, thus its quantum Fisher information increases... This explains the fluctuation of quantum Fisher information graphs, and it is the same analysis as that of fidelity graphs. Therefore, in the same reason, it is easily deduced why the width of fluctuation enlarges and the amplitude, where peaks and troughs appear, increases as $d$ grows.

We visualize the ratio between amplification schemes $\mathcal{R} = \mathcal{F}_{SCS}^{\hat{a}\hat{a}^\dagger}/\mathcal{F}_{SCS}^{\hat{a}^\dagger{}^2}$ for $d=2, 3, 4, 5$ in Fig. 5, as we did for hybrid states to compare quantum Fisher information of each scheme. The graph seems to coincide with our earlier qualitative analysis. However, we point out that some regions where $\hat{a}\hat{a}^\dagger$ yields larger quantum Fisher information than that of $\hat{a}^\dagger{}^2$ appear as $d$ increases. For example, for $\ket{\mathcal{C}^0_{\alpha,5}}$, $\mathcal{R}$ is larger than $1$ in $1.9 < \alpha <2.5$ region, where $\mathcal{F}_{SCS}^{\hat{a}^\dagger{}^2}$ is in trough. For sufficiently large amplitude, $\mathcal{R}$ converges to $1$, and the two schemes do not show a difference.

\begin{figure}[h]
    \centering
    \begin{subfigure}[h]{\linewidth}
        \centering
        \includegraphics[width=\textwidth]{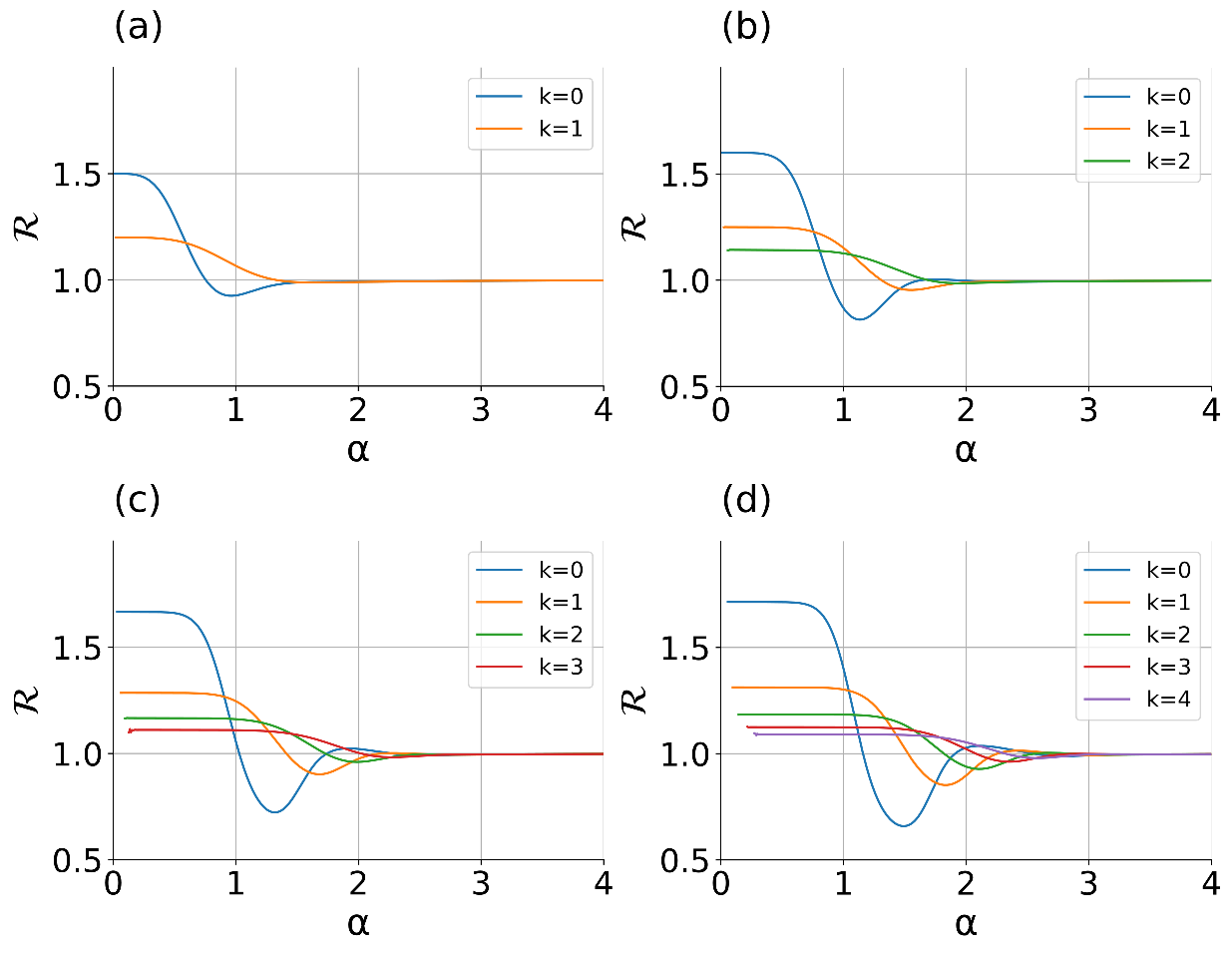}
    \end{subfigure}
    \caption{Ratio $\mathcal{R} = \mathcal{F}_{SCS}^{\hat{a}\hat{a}^\dagger}/\mathcal{F}_{SCS}^{\hat{a}^\dagger{}^2}$ against coherent state amplitude $\alpha$ for (a) $d=2$, (b) $d=3$, (c) $d=4$, and (d) $d=5$.}
    \label{SCS Fisher Information ratio}
\end{figure}

\begin{figure}[h]
    \centering
    \begin{subfigure}[h]{\linewidth}
        \centering
        \includegraphics[width=\textwidth]{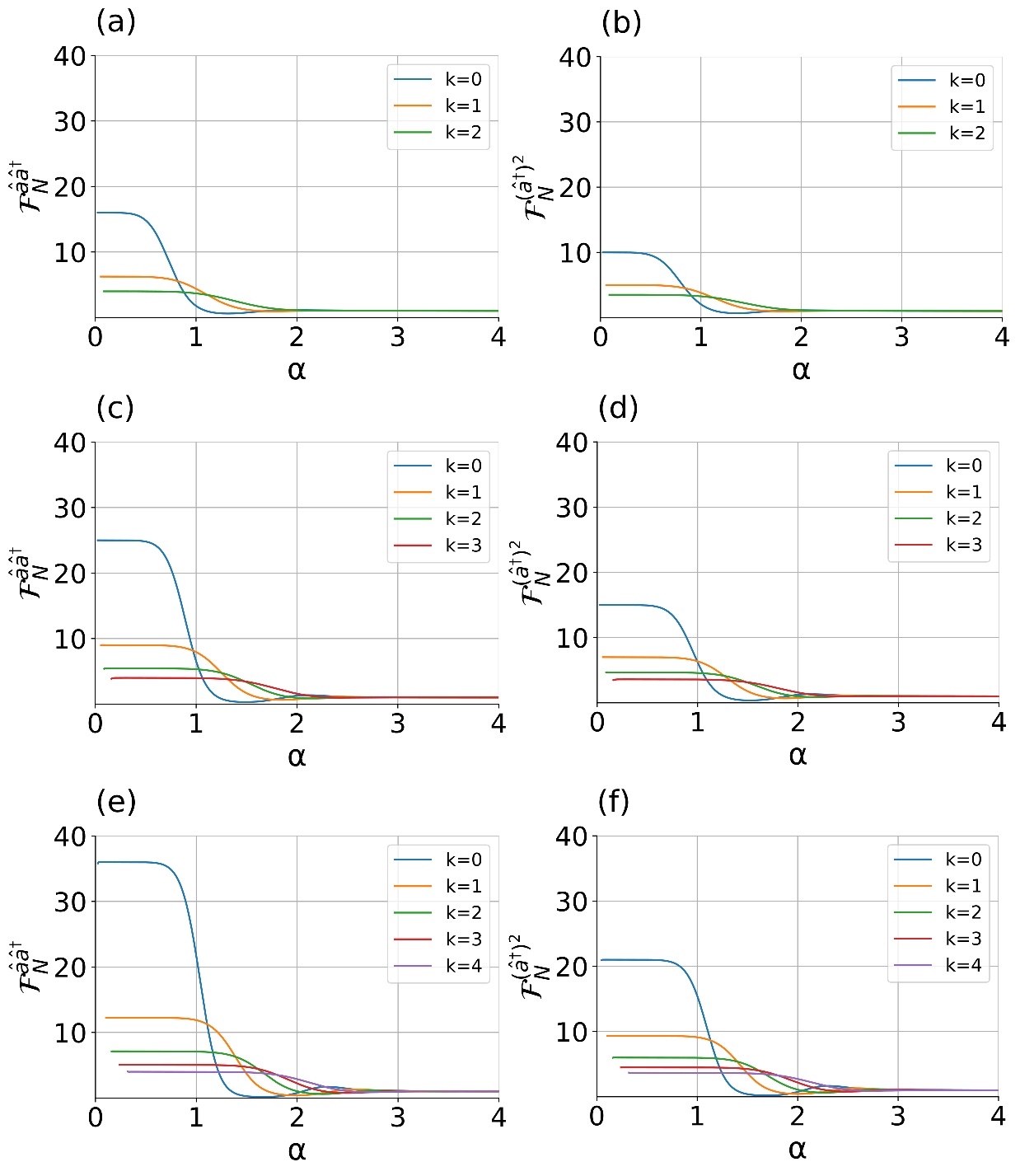}
    \end{subfigure}
    \caption{Normalized Fisher information $\mathcal{F}_N$ of SCSs against input coherent state amplitude $\alpha$, where (a) $d=3$ with $\hat{a}\hat{a}^\dagger$, (b) $d=3$ with $\hat{a}^\dagger{}^2$, (c) $d=4$ with $\hat{a}\hat{a}^\dagger$, (d) $d=4$ with $\hat{a}^\dagger{}^2$,  (e) $d=5$ with $\hat{a}\hat{a}^\dagger$, and (f) $d=5$ with $\hat{a}^\dagger{}^2$. Here, $\mathcal{F}_N^{\hat{A}}$ is defined as $\mathcal{F}^{\hat{A}}/\mathcal{F}$, where $\hat{A}$ denotes an amplification scheme, $\hat{a}\hat{a}^\dagger$ or $\hat{a}^\dagger{}^2$, while  $\mathcal{F}$ is quantum Fisher information of the input SCSs.}
    \label{SCS Fisher Information}
\end{figure}

The graphs of the ratio of quantum Fisher information of amplified SCSs over those of SCSs are plotted in Fig. 6, for $d=3, 4, 5$ with every $k$. For small amplitude, $\mathcal{F}_N$ of all states under the two schemes are much higher than $1$, which means that the performance of amplification for quantum phase estimation is good. After the plateau, the graphs rapidly decrease and fluctuate for SCSs with high $d$, and reach $1$ eventually. Due to the non-monotonic behavior of graphs, an amplification scheme should be chosen considering the goals of the experimentalist and the conditions of SCSs.

\section{Success probabilities of amplification schemes}

In this section, we analyze the success probabilities of both amplification schemes on HESs and SCSs. Our analysis in the previous sections is not restricted to specific methods of implementation, while the success probabilities depend on implementation schemes. Here, in order to provide some intuition regarding the success probabilities, we consider simple linear optical setups with single-photon sources and single-photon resolving detectors that distinguish between the vacuum and a single photon.

\begin{figure}[h]
    \centering
    \begin{subfigure}[h]{\linewidth}
        \centering
        \includegraphics[width=\textwidth]{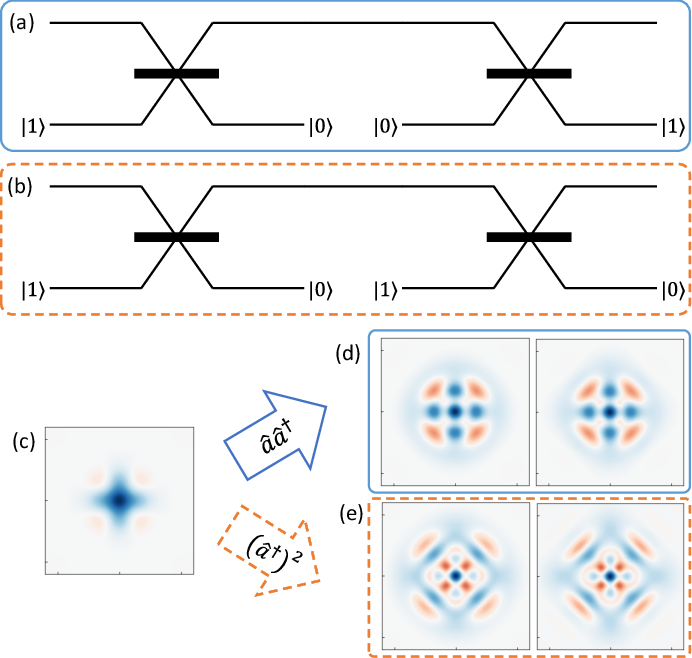}
    \end{subfigure}
    \caption{Schematic diagram of (a) $\hat{a}\hat{a}^\dagger{}$ scheme and (b) $\hat{a}^\dagger{}^2$ scheme. (c) Wigner function plot of an input SCS ($\ket{\mathcal{C}_{1,4}^{0}}$). The $\hat{a}\hat{a}^\dagger{}$ scheme realized with the BSs ($\gamma=0.01$) yields a state (d, left) $\ket{\mathcal{C}^{0, \hat{a}\hat{a}^\dagger{}}_{1, 4}}$ which is a good approximation of (d, right) SCS with larger amplitude, $\ket{\mathcal{C}_{1.5, 4}^{0}}$. The $\hat{a}^\dagger{}^2$ scheme of (b) also yields a state of (e, left) which is a good approximation of (e, right), $\ket{\mathcal{C}^{0, (\hat{a}^\dagger)^2}_{1, 4}}$, another SCS qudit with larger amplitude and different qudit number $k$, $\ket{\mathcal{C}_{2, 4}^{2}}$.}
    \label{amp_schematics}
\end{figure}

The amplification methods are depicted in Fig. \ref{amp_schematics}. Let us first consider the $\hat{a}\hat{a}^\dagger{}$ scheme in Fig. \ref{amp_schematics}(a). The single-photon addition may be performed using a single-photon state mixed with the input state at an almost transparent beam splitter followed by the vacuum measurement using a photodetector \cite{note1,Z2004}. The subsequent photon subtraction can be implemented by another beam splitter with high transmissivity to mix the input state with the vacuum and another photodetector that registers a single photon. The resulting state is then the photon-added-and-subtracted state on the input state. The $(\hat{a}^\dagger{})^2$ scheme may be performed in a similar way using two sequential photon addition processes as in Fig. \ref{amp_schematics}(b).

The state after the photon subtraction or addition can be understood as an element of the consequent ensemble of the corresponding quantum channel. Exact forms of related Kraus operators can be described as
\begin{equation}\label{opr_kraus}
\begin{aligned}
        \hat{K}_{-}|\psi\rangle_{s} &={}_{a}\langle1|\hat{U}_{BS}|\psi\rangle_{s}|0\rangle_a = \sqrt{\gamma}(1-\gamma)^{\frac{\hat{n}}{2}}\hat{a},\\
        \hat{K}_{+}|\psi\rangle_{s} &={}_{a}\langle0|\hat{U}_{BS}|\psi\rangle_{s}|1\rangle_a = \sqrt{\gamma}(1-\gamma)^{\frac{\hat{n}+1}{2}}\hat{a}^\dagger{},
\end{aligned}
\end{equation}
where $K_{-}(K_+)$ represents the Kraus operator corresponding to photon subtraction (addition). Here, $\gamma:=\mathrm{cos}^2\theta\ll1$ is the transmissivity of the beam splitter unitary $U_{BS}:=e^{i\theta(\hat{a}_s^\dagger{}\hat{a}_a + \hat{a}_s\hat{a}_a^\dagger{})}$, and the subscripts mark operators whether they are in system mode ($s$) or ancilla mode ($a$).

For small $\gamma$, the operation is approximately $\hat{a}$ $(\hat{a}^\dagger{})$. The success probability of each operation is now given as
\begin{equation}
\begin{aligned}
        P_-(|\psi\rangle) &= \mathrm{Tr}[\hat{K}_-^\dagger{}\hat{K}_-|\psi\rangle\langle\psi|], \\
        P_+(|\psi\rangle) &= \mathrm{Tr}[(\hat{K}_{+})^\dagger{}\hat{K}_{+}|\psi\rangle\langle\psi|]. \\
\end{aligned}
\end{equation}
The success probability of each amplification scheme, $\hat{a}\hat{a}^\dagger{}$ and $\hat{a}^\dagger{}^2$, can be analogously obtained from respective composite Kraus operators, $\hat{K}_{-}\hat{K}_{+}$ and $\hat{K}_+^2$ as
\begin{equation}
\begin{aligned}
        P_{\hat{a}\hat{a}^\dagger{}}(|\psi\rangle) &= \mathrm{Tr}[(\hat{K}_{-}\hat{K}_{+})^\dagger{}\hat{K}_{-}\hat{K}_{+}|\psi\rangle\langle\psi|], \\
        P_{\hat{a}^\dagger{}^2}(|\psi\rangle) &= \mathrm{Tr}[(\hat{K}_+^2)^\dagger{}\hat{K}_+^2|\psi\rangle\langle\psi|]. \\
\end{aligned}
\end{equation}

\begin{figure}
    \centering
    \begin{subfigure}[h]{\linewidth}
        \centering
    `   \includegraphics[width=\textwidth]{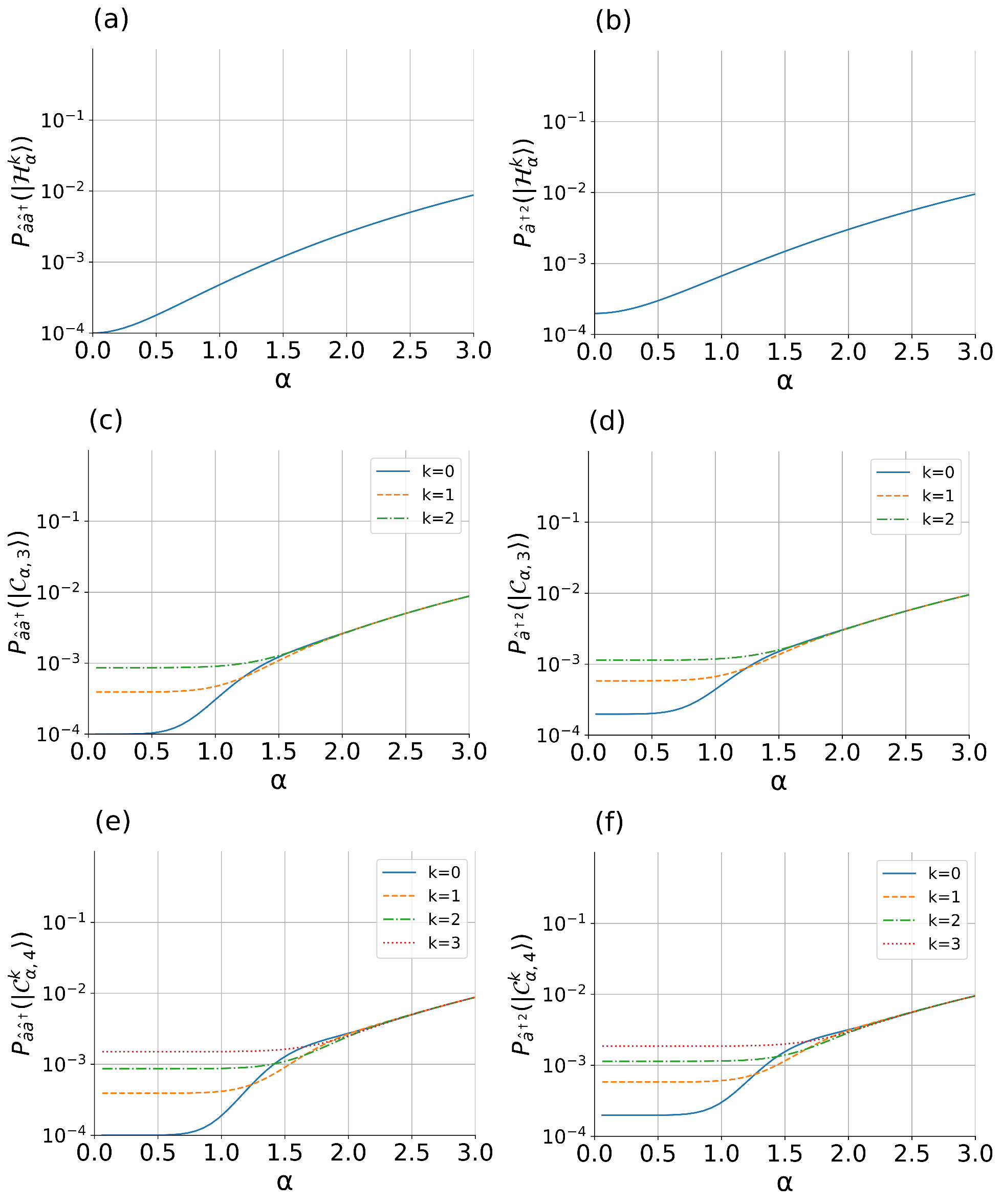}
    \end{subfigure}
    \caption{Success probabilities of the amplification schemes acting on HESs and SCSs using beam splitters with $\gamma=0.01$. Amplification success probabilities of HESs using (a) $\hat{a}\hat{a}^\dagger{}$ scheme and (b) $\hat{a}^\dagger{}^2$ scheme. Qudit number is not specified since the result is the same for all $k$. Amplification success probabilities of SCSs where (c) $d=3$ with $\hat{a}\hat{a}^\dagger{}$, (d) $d=3$ with $\hat{a}^\dagger{}^2$, (e) $d=4$ with $\hat{a}\hat{a}^\dagger{}$, and (f) $d=4$ with $\hat{a}^\dagger{}^2$.}
    \label{success_prob}

\end{figure}

Figure~\ref{success_prob} shows the success probabilities of linear optical amplification schemes with $\gamma=0.01$ acting on HESs and SCSs. Data were obtained in two ways: numerical simulation of the linear optical circuit using a truncated Fock basis (for $n\leq 30$) and estimation of probabilities using the Kraus operators discussed above. The results of both schemes (numerical simulation and analytic calculation) are in agreement with each other. The HESs again show uniform behavior regardless of the qudit number $k$, since it is locally equivalent to a classical mixture of coherent states with the same amplitude. 

The SCSs show more complicated behaviors. For large $\alpha$, the qudit number $k$ does not effectively affect the success probability, just as the cases for other quantities analyzed in the previous sections (fidelity, optimized gain, and QFI). 
For small amplitudes, however, the SCSs are approximated as Fock states as $|\mathcal{C}^k_{\alpha, d}\rangle \approx |k\rangle$. In this regime, the effects of the photon addition and the subtraction are clearly distinct depending on $k$, and 
the change of the qudit number $k$ clearly changes the average photon number of the state. This is not the case for large $\alpha$ where $|\mathcal{C}^k_{\alpha, d}\rangle$'s have nearly the same average photon number for given $\alpha$ regardless of the value of $k$.
In fact, the leading order term of both the Kraus operators ($\hat{K}_+$ and $\hat{K}_-$) for small gamma is $\gamma\hat{n}$.
Therefore, the changes of $k$ considerably affects the success probability only for small $\alpha$.
The leading term of the Kraus operators also implies the uniform increase in the larger amplitude regime by increasing $\alpha$ regardless of the qudit number and the amplification scheme.

We can thus generally point out that a larger success probability can be achieved by an input state with a larger amplitude, while it has a trade-off of smaller gain. When $\alpha$ is small, it is better to choose a large value of $k$ in order to increase the success probability.
High success probabilities may also be achieved by increasing the beam-splitter transmissivity $\gamma$, but it would decrease the fidelity.

\section{Conclusions}

We have investigated two amplification schemes, the photon-addition-and-then-subtraction scheme ($\hat{a}^\dag \hat{a}$) and the successive-photon-addition scheme ($\hat{a}^\dagger{}^2$), applied to two types of nonclassical states, {\it i.e.,} HESs and SCSs. A difference between HESs and SCSs is that the former is a two-mode state with entanglement while the latter is a single-mode state. A HES is locally indistinguishable from a classical mixture of two coherent states if only the coherent-state mode is considered. Thus, it is natural that the results of the amplification schemes with a HES are similar to those with a coherent state of the same amplitude. However, the SCSs are single-mode superpositions of two coherent states and their behaviors are different. In a small amplitude regime, the SCSs are similar to the Fock states with a photon number same as their qudit number. With larger amplitudes, however, the SCSs are rather closer to coherent states with the same amplitudes except for the additional modular photon number condition. 

We have obtained the fidelity, the optimized gain, and the quantum Fisher information of each scheme for each state. For both states, the $\hat{a}\hat{a}^\dagger{}$ scheme has the advantage of higher fidelity, and it is appropriate for amplifying states with relatively small amplitudes ($\alpha\lesssim 1.5$). On the contrary, the $\hat{a}^\dagger{}^2$ scheme has the advantage of higher gain, and it is better for amplifying states with intermediate amplitudes ($1.5\lesssim\alpha\lesssim 2.5$). For sufficiently large amplitudes, the difference between the two schemes vanishes.

For the HESs, the fidelity and gain under both schemes are the same as those for the case of coherent states, independent of dimension $d$ or qudit number $k$ due to the orthogonality of the DV part. We have obtained the quantum Fisher information of the HESs that reflects the overall performance of the noiseless amplification. Our analysis shows that the $\hat{a}^\dag \hat{a}$ scheme is better for $\alpha\lesssim 0.9$, while the $\hat{a}^\dagger{}^2$ scheme works generally better for $\alpha$ larger than $0.9$.

For the SCSs, the behaviors of the fidelity and gain for the amplified states depend on dimension $d$ and qudit number $k$. In the fidelity graphs, a smaller qudit number $k$ shows trough(s) for a smaller amplitude with larger depth. In the gain graphs, similar behaviors are observed but two states with the last two qudit numbers $d-2$ and $d-1$ act very differently. We have analyzed the graphs focusing on the pseudo-number property of the superposition of coherent states. Unlike the HESs, the quantum Fisher information graphs show fluctuations as the amplitude increases and thus do not reflect the overall performance of noiseless amplification except $d=2$. Consequently, the quantum Fisher information is adopted as an example of the performance of the amplified SCS, and consistent criterion extended from even and odd cat states, to compare with other optical states. Since the fidelity and gain graphs are not trivial, and the quantum Fisher information does not indicate overall performance, an appropriate amplification scheme should be chosen considering the goals and conditions of initial states.

 \section*{Acknowledgements}
The first two authors contributed equally to this work. This work was supported by the National Research Foundation of Korea (NRF) grant funded by the Korea government (MSIT) (Nos. RS-2024-00413957, RS-2024-00438415, NRF-2023R1A2C1006115, and RS-2024-00437191) and by the Institute of Information \& Communications Technology Planning \& Evaluation (IITP) grant funded by the Korea government (MSIT) (IITP-2024-2020-0-01606). This work was also supported by the Institute of Applied Physics, Seoul National University.

\section{Appendices}
\subsection{Appendix A}
The fidelity, the gain, and the quantum Fisher information of hybrid states are calculated here. The fidelity between the target state and the amplified $d$-dimensional hybrid states under the photon addition and then subtraction scheme is
\begin{align} \nonumber
      F^{\hat{a}\hat{a}^\dagger}_{HES} &= N_{HES}^{\hat{a}\hat{a}^\dagger}(\alpha,d)^2  |\frac{1}{d}\sum_{n,m} \bra{n, g \alpha \omega_d^n} \hat{a}\hat{a}^\dagger \ket{m, \alpha \omega_d^m}|^2 \\ \nonumber
       &= \frac{|\sum_n \bra{g \alpha \omega_d^n} (1+ \hat{a}^\dagger \hat{a}) \ket{\alpha \omega_d^n}|^2 }{ d^2 (\alpha^4 + 3\alpha^2 +1) } \\
       &= \frac{g^2 \alpha^4 + 2 g \alpha^2 +1}{ \alpha^4 + 3\alpha^2 +1 } \mathrm{exp}[-\alpha^2(g-1)^2].
   \end{align}
The fidelity between the target state and the amplified state for $d$-dimensional hybrid states under the $(\hat{a}^\dag)^2$ scheme is
  \begin{align}  \nonumber
      F_{HES}^{\hat{a}^\dagger{}^2} &= N_{HES}^{\hat{a}^\dagger{}^2}(\alpha,d)^2  |\frac{1}{d}\sum_{n,m} \omega_d^{2n} \bra{n, g \alpha \omega_d^n} \hat{a}^\dagger{}^2 \ket{m, \alpha \omega_d^m}|^2 \\ \nonumber
       &= \frac{|\sum_n \omega_d^{2n} \bra{g \alpha \omega_d^n} (\hat{a}^\dagger{}^2) \ket{\alpha \omega_d^n}|^2 }{ d^2 (\alpha^4 + 4\alpha^2 +2) } \\
       &= \frac{g^4 \alpha^4 }{ (\alpha^4 + 4\alpha^2 +2) } \mathrm{exp}[-\alpha^2(g-1)^2].
  \end{align}

The gain values that maximize the fidelity are obtained by solving the Eq. (\ref{def of gain}). For $\hat{a}\hat{a}^\dagger$ scheme, the equation reduces to
   \begin{align}
      (G^{\hat{a}\hat{a}^\dagger}_{HES} \alpha^2+1)\{1- (G^{\hat{a}\hat{a}^\dagger}_{HES} \alpha^2+1)(G^{\hat{a}\hat{a}^\dagger}_{HES}-1) \} =0.
   \end{align}
Since $g$ and $\alpha$ are positive, we have
   \begin{align}
       1- (G^{\hat{a}\hat{a}^\dagger}_{HES} \alpha^2+1)(G^{\hat{a}\hat{a}^\dagger}_{HES}-1) =0,
   \end{align}
which gives
   \begin{align}
       G^{\hat{a}\hat{a}^\dagger}_{HES} = \frac{\alpha^2 -1 + \sqrt{\alpha^4 + 6 \alpha^2 +1}}{2 \alpha^2}.
   \end{align}  
For $(\hat{a}^\dag)^2$ scheme, the Eq. (\ref{def of gain}) reduces to
   \begin{align}
       (G^{\hat{a}^\dagger{}^2}_{HES})^2 - G^{\hat{a}^\dagger{}^2}_{HES} - \frac{2}{\alpha^2} = 0,
   \end{align}
which gives
   \begin{align}
       G^{\hat{a}^\dagger{}^2}_{HES} = \frac{1 + \sqrt{1+\frac{8}{\alpha^2}}}{2}.
   \end{align}

The quantum Fisher information of a pure state concerning the observable $\hat{H}$ is given by $4\langle \Delta \hat{H}\rangle$ which is proportional to $4\langle \Delta \hat{n}\rangle$ for optics system. Using creation operators and annihilation operators and their relation $[\hat{a},\hat{a}^\dagger]=1$, we will compute $4\langle \hat{a}^\dagger{}^2 \hat{a}^2 \rangle$ + $4\langle \hat{a}^\dagger \hat{a} \rangle$ - $4\langle \hat{a}^\dagger \hat{a} \rangle^2$. The quantum Fisher information of bare hybrid states is calculated as
    \begin{align} \nonumber
           &\mathcal{F}_{HES}= \frac{4}{d} \sum_{n} \bra{\alpha \omega_d^n} (\hat{a}^\dagger)^2 \hat{a}^2 \ket{\alpha \omega_d^n}\\\nonumber
            + &\frac{4}{d} \sum_{n} \bra{\alpha \omega_d^n} \hat{a}^\dagger \hat{a} \ket{\alpha     \omega_d^n} - 4(\frac{1}{d} \sum_{n} \bra{\alpha \omega_d^n} \hat{a}^\dagger \hat{a} \ket{\alpha \omega_d^n})^2 \\
        = &4\alpha^4 + 4\alpha^2 - 4\alpha^4 = 4\alpha^2,
    \end{align}
which is the same result with coherent states. Second, the quantum Fisher information of amplified hybrid states under the photon addition and then subtraction scheme is
    \begin{align}\nonumber
           &\mathcal{F}_{HES}^{\hat{a}\hat{a}{}^\dagger} = 4 {(N_{HES}^{\hat{a}\hat{a}{}^\dagger}})^2 ( \frac{1}{d} \sum_{n} \bra{\alpha \omega_d^n} \hat{a}\hat{a}^\dagger (\hat{a}^\dagger)^2 \hat{a}^2 \hat{a}\hat{a}^\dagger\ket{\alpha \omega_d^n} \\ \nonumber
           & + \frac{1}{d} \sum_{n} \bra{\alpha \omega_d^n} \hat{a}\hat{a}^\dagger \hat{a}^\dagger \hat{a} \hat{a}\hat{a}^\dagger \ket{\alpha \omega_d^n}) \\
           & - 4({N_{HES}^{\hat{a}\hat{a}^\dagger}})^4 (\frac{1}{d} \sum_{n} \bra{\alpha \omega_d^n} \hat{a}\hat{a}^\dagger \hat{a}^\dagger \hat{a} \hat{a}\hat{a}^\dagger \ket{\alpha \omega_d^n})^2.
    \end{align}
    
To calculate this, observe that
    \begin{align}
        \hat{a} (\hat{a}^\dagger)^3 \hat{a}^3 \hat{a}^\dagger &= (\hat{a}^\dagger)^4\hat{a}^4 + 7(\hat{a}^\dagger)^3\hat{a}^3 + 9(\hat{a}^\dagger)^2\hat{a}^2, \\
        \hat{a} (\hat{a}^\dagger)^2 \hat{a}^2 \hat{a}^\dagger &= (\hat{a}^\dagger)^3\hat{a}^3 + 5(\hat{a}^\dagger)^2\hat{a}^2 +4(\hat{a}^\dagger)\hat{a}.
    \end{align}
    
Therefore, the quantum Fisher information is calculated to be
    \begin{align}
        \mathcal{F}_{HES}^{\hat{a}\hat{a}^\dagger} = \frac{4\alpha^2(\alpha^8+6\alpha^6+14\alpha^4+10\alpha^2+4)}{(\alpha^4+3\alpha^2+1)^2}.
    \end{align}
    
Third, after $(\hat{a}^\dag)^2$ to hybrid states, the Fisher information is given by
     \begin{align}\nonumber
         &\mathcal{F}_{HES}^{\hat{a}^\dagger{}^2} = 4({N_{HES}^{(\hat{a}^\dagger{})^2}})^2 ( \frac{1}{d} \sum_{n} \bra{\alpha \omega_d^n} \hat{a}^2 (\hat{a}^\dagger)^2 \hat{a}^2 (\hat{a}^\dagger)^2\ket{\alpha \omega_d^n} \\ \nonumber
         & + \frac{1}{d} \sum_{n} \bra{\alpha \omega_d^n} \hat{a}^2 \hat{a}^\dagger \hat{a} (\hat{a}^\dagger)^2 \ket{\alpha \omega_d^n}) \\
         & - 4 ({N_{HES}^{(\hat{a}^\dagger{})^2}})^2 (\frac{1}{d} \sum_{n} \bra{\alpha \omega_d^n} \hat{a}^2 \hat{a}^\dagger \hat{a} (\hat{a}^\dagger)^2 \ket{\alpha \omega_d^n})^2.
     \end{align}
To calculate this, observe that
    \begin{align} \nonumber
         &\hat{a}^2 (\hat{a}^\dagger)^2 \hat{a}^2 (\hat{a}^\dagger)^2 \\
         &= (\hat{a}^\dagger)^4\hat{a}^4 + 12(\hat{a}^\dagger)^3\hat{a}^3 + 38(\hat{a}^\dagger)^2\hat{a}^2 + 32(\hat{a}^\dagger)\hat{a} + 4, \\
         &\hat{a}^2 (\hat{a}^\dagger) \hat{a} (\hat{a}^\dagger)^2 = (\hat{a}^\dagger)^3\hat{a}^3 + 8(\hat{a}^\dagger)^2\hat{a}^2 +14(\hat{a}^\dagger)\hat{a} +4.
    \end{align}
Therefore, Eq. (54) is calculated to be
   \begin{align}
       \mathcal{F}_{HES}^{\hat{a}^\dagger{}^2} = \frac{4\alpha^2(\alpha^8+8\alpha^6+24\alpha^4+24\alpha^2+12)}{(\alpha^4+4\alpha^2+2)^2}.
   \end{align}

\subsection{Appendix B}
Here we compute the fidelity and the quantum Fisher information of the SCSs. First, we calculate the fidelity between the target state and the amplified $d$-dimensional SCSs with qudit number $k$ under the photon addition and then subtraction scheme, which is denoted by $ F_{SCS}^{\hat{a}\hat{a}^\dagger}$, and under the $(\hat{a}^\dag)^2$ scheme, which is denoted by $ F_{SCS}^{\hat{a}^\dagger{}^2}$, in order. We obtain expressions as follows:
   \begin{widetext}
        \begin{align}\nonumber
             F_{SCS}^{\hat{a}\hat{a}^\dagger} &= N_{SCS}^{\hat{a}\hat{a}^\dagger}(\alpha,k,d)^2 N_{k,g \alpha,d}^2 |\sum_{m,n} \bra{g \alpha \omega_d^m} \omega_d^{km}\hat{a}\hat{a}^\dagger \omega_d^{-kn} \ket{\alpha \omega_d^n}|^2 \\ \nonumber
             &= \frac{| \sum_{m,n} (1+g\alpha^2 \omega_d^{(n-m)})\omega_d^{-k(n-m)} \bra{g \alpha \omega_d^m} \ket{\alpha \omega_d^n}|^2 }{ d^2 (\sum_n \omega_d^{-kn}(\alpha^4 \omega_d^{2n} + 3\alpha^2 \omega_d^n+1)\mathrm{exp}[-\alpha^2(1-\omega_d^n)])(\sum_n \omega_d^{-kn}\mathrm{exp}[-g^2\alpha^2(1-\omega_d^n)]) } \\
             &= \frac{\{\sum_n (1+g\alpha^2\omega_d^{n}) \omega_d^{-kn} \mathrm{exp}[g\alpha^2 \omega_d^n] \}^2}{(\sum_n \omega_d^{-kn}(\alpha^4 \omega_d^{2n} + 3\alpha^2\omega_d^n+1)\mathrm{exp}[\alpha^2\omega_d^n])(\sum_n \omega_d^{-kn}\mathrm{exp}[g^2\alpha^2\omega_d^n]) },
         \end{align}
         \begin{align}\nonumber
             F_{SCS}^{\hat{a}^\dagger{}^2} &= N_{SCS}^{\hat{a}^\dagger{}^2}(\alpha,k,d)^2 N_{k+2,g \alpha,d}^2 |\sum_{m,n} \bra{g \alpha \omega_d^m} \omega_d^{(k+2)m}\hat{a}^\dagger{}^2 \omega_d^{-kn} \ket{\alpha \omega_d^n}|^2 \\ \nonumber
             &= \frac{| g^2 \alpha^2 \sum_{m,n} \omega_d^{-k (n-m)} \bra{g \alpha \omega_d^m} \ket{\alpha \omega_d^n}|^2 }{ d^2 (\sum_n \omega_d^{-kn}(\alpha^4 \omega_d^{2 n} + 4\alpha^2 \omega_d^n+2)\mathrm{exp}[-\alpha^2(1-\omega_d^n)])(\sum_n \omega_d^{-(k+2)n}\mathrm{exp}[-g^2\alpha^2(1-\omega_d^n)]) } \\
             &= \frac{g^4\alpha^4 \{ \sum_n \omega_d^{- k n} \mathrm{exp}[g\alpha^2 \omega_d^n] \}^2}{(\sum_n \omega_d^{-kn}(\alpha^4 \omega_d^{2n} + 4\alpha^2\omega_d^n +2)\mathrm{exp}[\alpha^2\omega_d^n])(\sum_n \omega_d^{-(k+2)n}\mathrm{exp}[g^2\alpha^2\omega_d^n]) }.
         \end{align}
\end{widetext}
These expressions are too complicated to apply and solve Eq. (\ref{def of gain}). Thus, we do not obtain the gain in closed forms. Instead, we find numerical values of the gain using the downhill simplex algorithm through Python package Scipy, using the ~\textbf{scipy fmin} function.

Second, we calculate the quantum Fisher information of a superposition of coherent states with dimension $d$ and qudit number $k$, which is denoted by $\mathcal{F}_{SCS}$, that of amplified superposition of coherent states with dimension $d$ and qudit number $k$ under the $\hat{a}\hat{a}^\dagger$ scheme, $\mathcal{F}_{SCS}^{\hat{a}\hat{a}^\dagger}$, and under the $\hat{a}^\dagger{}^2$ scheme, $\mathcal{F}_{SCS}^{\hat{a}^\dagger{}^2}$, in order.
\begin{widetext}
        \begin{align}\nonumber
            &\mathcal{F}_{SCS} = 4\, (N_{k,\alpha,d})^2(\frac{1}{d} \sum_{m,n} \bra{\alpha \omega_d^m} \omega_d^{km}(\hat{a}^\dagger)^2 \hat{a}^2 \omega_d^{-kn}\ket{\alpha \omega_d^n}+ \frac{1}{d} \sum_{m,n} \bra{\alpha \omega_d^m} \omega_d^{km} \hat{a}^\dagger \hat{a} \omega_d^{-kn} \ket{\alpha \omega_d^n})  \\ \nonumber
            &- 4\,((N_{k,\alpha,d})^2 \frac{1}{d} \sum_{m,n} \bra{\alpha \omega_d^m} \omega_d^{km} \hat{a}^\dagger \hat{a} \omega_d^{-kn}\ket{\alpha \omega_d^n})^2= 4 (N_{k,\alpha,d})^2 \sum_{m,n} (\alpha^4 \omega_d^{2(n-m)} +\alpha^2 \omega_d^{(n-m)}) \omega_d^{-k(n-m)} \mathrm{exp}[-\alpha(1-\omega_d^{(n-m)})] \\ \nonumber
            &-4 \,(N_{k,\alpha,d})^4(\sum_{m,n}\alpha^2 \omega_d^{(n-m)} \omega_d^{-k(n-m)} \mathrm{exp}[-\alpha^2(1-\omega_d^{(n-m)})])^2 \\
            &= 4 \alpha^2 \frac{ \sum_n (\alpha^2 \omega_d^{(2-k)n} + \omega_d^{(1-k)n}) \mathrm{exp}[\alpha^2\omega_d^n]}{ \sum_n \omega_d^{-kn} \mathrm{exp}[\alpha^2\omega_d^n]} - 4 \alpha^4 \left(\frac{\sum_n \omega_d^{(1-k)n} \mathrm{exp}[\alpha^2\omega_d^n]}{ \sum_n \omega_d^{-kn} \mathrm{exp}[\alpha^2\omega_d^n]}\right)^2,
        \end{align}

       \begin{align}\nonumber
            &\mathcal{F}_{SCS}^{\hat{a}\hat{a}^\dagger} = 4\,( {N_{SCS}^{\hat{a}\hat{a}^\dagger}})^2 (\frac{1}{d} \sum_{m,n} \bra{\alpha \omega_d^m}  \omega_d^{km}\hat{a}\hat{a}^\dagger (\hat{a}^\dagger)^2 \hat{a}^2 \hat{a}\hat{a}^\dagger \omega_d^{-kn}\ket{\alpha \omega_d^n}+ \frac{1}{d} \sum_{m,n} \bra{\alpha \omega_d^m} \omega_d^{km} \hat{a}\hat{a}^\dagger \hat{a}^\dagger \hat{a} \hat{a}\hat{a}^\dagger \omega_d^{-kn} \ket{\alpha \omega_d^n}) \\ \nonumber
            &-4\, ((N_{k,\alpha,d})^2 \frac{1}{d} \sum_{m,n} \bra{\alpha \omega_d^m} \omega_d^{km} \hat{a}\hat{a}^\dagger \hat{a}^\dagger \hat{a} \hat{a}\hat{a}^\dagger \omega_d^{-kn}\ket{\alpha \omega_d^n})^2 \\ \nonumber
            &= 4\,({N_{SCS}^{\hat{a}\hat{a}^\dagger}})^2 \sum_{m,n} (\alpha^8 \omega_d^{4(n-m)} +8 \alpha^6 \omega_d^{3(n-m)} +14 \alpha^4 \omega_d^{2(n-m)} + 4 \alpha^2 \omega_d^{(n-m)})\omega_d^{-k(n-m)} \mathrm{exp}[-\alpha(1-\omega_d^{(n-m)})] \\ \nonumber
            &- 4\,({N_{SCS}^{\hat{a}\hat{a}^\dagger}})^4 \{\sum_{m,n} (\alpha^6 \omega_d^{3(n-m)} +5 \alpha^4 \omega_d^{2(n-m)} + 4 \alpha^2 \omega_d^{(n-m)})\omega_d^{-k(n-m)} \mathrm{exp}[-\alpha(1-\omega_d^{(n-m)})]\}^2 \\
            &= 4\,\frac{ \sum_n (\alpha^8 \omega_d^{4n} +8 \alpha^6 \omega_d^{3n} +14 \alpha^4 \omega_d^{2n} + 4 \alpha^2 \omega_d^{n}) \mathrm{exp}[\alpha^2\omega_d^n] \omega_d^{-kn}}{ \sum_n (\alpha^4 \omega_d^{2n} + 3 \alpha^2 \omega_d^{n} + 1 )\omega_d^{-kn} \mathrm{exp}[\alpha^2\omega_d^n]} -  \left( \frac{ \sum_n (\alpha^6 \omega_d^{3n} +5 \alpha^4 \omega_d^{2n} + 4 \alpha^2 \omega_d^{n}) \mathrm{exp}[\alpha^2\omega_d^n] \omega_d^{-kn}}{ \sum_n (\alpha^4 \omega_d^{2n} + 3 \alpha^2 \omega_d^{n} +1 )\omega_d^{-kn} \mathrm{exp}[\alpha^2\omega_d^n]} \right)^2,
       \end{align}
       \begin{align} \nonumber
            &\mathcal{F}_{SCS}^{\hat{a}^\dagger{}^2} = 4\,({N_{SCS}^{(\hat{a}^\dagger{})^2}}) ^2 (\frac{1}{d} \sum_{m,n} \bra{\alpha \omega_d^m}  \omega_d^{km}\hat{a}^2 (\hat{a}^\dagger)^2 \hat{a}^2 \hat{a}^\dagger{}^2 \omega_d^{-kn}\ket{\alpha \omega_d^n}+ \frac{1}{d} \sum_{m,n} \bra{\alpha \omega_d^m} \omega_d^{km} \hat{a}^2\hat{a}^\dagger \hat{a} \hat{a}^\dagger{}^2 \omega_d^{-kn} \ket{\alpha \omega_d^n}) \\ \nonumber
            &- 4\, ((N_{k,\alpha,d})^2 \frac{1}{d} \sum_{m,n} \bra{\alpha \omega_d^m} \omega_d^{km} \hat{a}^2 \hat{a}^\dagger \hat{a} \hat{a}^\dagger{}^2 \omega_d^{-kn}\ket{\alpha \omega_d^n})^2 \\ \nonumber
            &= 4\,{(N_{SCS}^{(\hat{a}^\dagger{})^2}})^2 \sum_{m,n} (\alpha^8 \omega_d^{4(n-m)} +13 \alpha^6 \omega_d^{3(n-m)} +46 \alpha^4 \omega_d^{2(n-m)} + 46 \alpha^2 \omega_d^{(n-m)} +8) \mathrm{exp}[-\alpha(1-\omega_d^{(n-m)})]\omega_d^{-k(n-m)} \\ \nonumber
            &- 4\,({N_{SCS}^{(\hat{a}^\dagger{})^2}})^4 \{\sum_{m,n} (\alpha^6 \omega_d^{3(n-m)}) +8 \alpha^4 \omega_d^{2(n-m)} + 14 \alpha^2 \omega_d^{(n-m)} +4)\omega_d^{-k(n-m)} \mathrm{exp}[-\alpha(1-\omega_d^{(n-m)})]\}^2 \\ \nonumber
            &= 4\,\frac{ \sum_n (\alpha^8 \omega_d^{4n} +13 \alpha^6 \omega_d^{3n} +46 \alpha^4 \omega_d^{2n} + 46 \alpha^2 \omega_d^{n} +8) \mathrm{exp}[\alpha^2\omega_d^n] \omega_d^{-kn}}{ \sum_n (\alpha^4 \omega_d^{2n} + 4 \alpha^2 \omega_d^{n} +2 )\omega_d^{-kn} \mathrm{exp}[\alpha^2\omega_d^n]} \\
            &- 4\,\left( \frac{ \sum_n (\alpha^6 \omega_d^{3n} +8 \alpha^4 \omega_d^{2n} + 14 \alpha^2 \omega_d^{n} +4) \mathrm{exp}[\alpha^2\omega_d^n] \omega_d^{-kn}}{ \sum_n (\alpha^4 \omega_d^{2n} + 4 \alpha^2 \omega_d^{n} +2 )\omega_d^{-kn} \mathrm{exp}[\alpha^2\omega_d^n]} \right)^2.
        \end{align}
    \end{widetext}


\begin{thebibliography}{26}


\bibitem{Jeong14} H. Jeong, A. Zavatta, M. Kang, S.-W Lee, L. S. Constanzo, S. Grandi, T. C. Ralph and M. Bellini, Generation of hybrid entanglement of light, Nat. Photonics~{\bf 8}, 564 (2014).
\bibitem{Morin14} O. Morin, K. Juang, J. Liu, H. L. Jeannic, C. Fabre, and J. Laurat, Remote creation of hybrid entanglement between particle-like and wave-like optical qubits, Nat. Photonics~{\bf 8}, 570 (2014).

\bibitem{Yurke86} B. Yurke and D. Stoler, Generating quantum mechanical superpositions of macroscopically distinguishable states via amplitude dispersion, Phys. Rev. Lett.~{\bf 57}, 13 (1986).
\bibitem{Schleich91} W. Schleich, M. Pernigo, and F. L. Kien, Nonclassical state from two pseudoclassical states, Phys. Rev. A~{\bf 44}, 2172 (1991).

\bibitem{Jeong23} H. Jeong, Converting qubits, Nat. Photonics ~{\bf 17}(2), 131 (2023)

\bibitem{Ulanov17} A.E. Ulanov, D. Sychev, A.A. Pushkina, I.A. Fedorov, and A.I. Lvovsky, Quantum teleportation between discrete and continuous encodings of an optical qubit, Phys. Rev. Lett. ~{\bf 118}, 160501 (2017).
\bibitem{Sychev18} D.V. Sychev, A.E. Ulanov, E.S. Tiunov, A.A. Pushkina, Entanglement and teleportation between polarization and wave-like encodings of an optical qubit, Nat. Commun. ~{\bf 9}, 3672 (2018).
\bibitem{Darras23} T. Darras, B. E. Asenbeck, G. Guccione, A. Cavaill{\'e}s, H. Le Jeannic, and J. Laurat, A quantum-bit encoding converter, Nat. Photonics, ~{\bf 17}(2), 165 (2023)


\bibitem{Kwon13} H. Kwon and H. Jeong, Violation of the Bell–Clauser-Horne-Shimony-Holt inequality using imperfect photodetectors with optical hybrid states, Phys. Rev. A ~{\bf 88}, 052127 (2013).
\bibitem{Park12} K. Park, S.-W Lee, and H. Jeong, Quantum teleportation between particlelike and fieldlike qubits using hybrid entanglement under decoherence effects, Phys. Rev. A ~{\bf 86}, 062301 (2012).
\bibitem{Jeong16} H. Jeong, S. Bae, and S. Choi, Quantum teleportation between a single-rail single-photon qubit and a coherent-state qubit using hybrid entanglement under decoherence effects, Quant. Info. Proc.~{\bf 15}, 913 (2016).
\bibitem{Kim16} H. Kim, S.-W Lee, and H. Jeong, Two different types of optical hybrid qubits for teleportation in a lossy environment, Quant. Info. Proc.~{\bf 15}, 4729 (2016).
\bibitem{Lee13} S.-W Lee, H. Jeong, Near-deterministic quantum teleportation and resource-efficient quantum computation using linear optics and hybrid qubits, Phys. Rev. A~{\bf 87}, 022326 (2013).

\bibitem{Omkar19} S. Omkar, Y.-S. Teo, and H. Jeong, Resource-efficient topological fault-tolerant quantum computation with hybrid entanglement of light, Phys. Rev. Lett. ~{\bf 125}, 060501 (2020).


\bibitem{Lim16} Y. Lim, J. Joo, T. P. Spiller, and H. Jeong, Loss-resilient photonic entanglement swapping using optical hybrid states, Phys. Rev. A ~{\bf 94}, 062337 (2016).



\bibitem{Bose23} S. Bose, J. Singh, A. Cabello, and H. Jeong, Long-distance entanglement sharing using hybrid states of discrete and continuous variables, Phys. Rev. Applied ~{\bf 21}, 064103 (2024).

\bibitem{JK2001} H. Jeong and M. S. Kim, Purification of entangled coherent states, Quantum Information and Computation {\bf 2}, 208 (2002).

\bibitem{Cavailles18} A. Cavaill\' es, H. Le Jeannic, J. Raskop, G. Guccione, D. Markham, E. Diamanti, M. D. Shaw, V. B. Verma, S. W. Nam, and J. Laurat, Demonstration of Einstein-Podolsky-Rosen steering using hybrid continuous-and discrete-variable entanglement of light, Phys. Rev. Lett. ~{\bf 121}, 170403 (2018).

\bibitem{Huang19} X. Huang, E. Zeuthen, Q. Gong, and Q. He, Engineering asymmetric steady-state Einstein-Podolsky-Rosen steering in macroscopic hybrid systems, Phys. Rev. A ~{\bf 100}, 012318 (2019).

\bibitem{Gerry99}  C. Gerry, Generation of optical macroscopic quantum superposition states via state reduction with a Mach-Zehnder interferometer containing a Kerr medium, Phys. Rev. A~{\bf 59}, 4095-4098 (1999).
\bibitem{Nemoto04}  K. Nemoto and W. J. Munro, Nearly deterministic linear optical controlled-NOT gate, Phys. Rev. Lett.~{\bf 93}, 250502 (2004).
\bibitem{Jeong05a}  H. Jeong, Using weak nonlinearity under decoherence for macroscopic entanglement generation and quantum computation, Phys. Rev. A ~{\bf 072} 034305, (2005).
\bibitem{Anderson13}  U. L. Anderson and J. S. Neergaard-Nielsen, Heralded generation of a micro-macro entangled state, Phys. Rev. A~{\bf 88}, 022337 (2013).




\bibitem{Cochrane99} P. T. Cochrane, G. J. Milburn, and W. J. Munro, Macroscopically distinct quantum-superposition states as a bosonic code for amplitude damping, Phys. Rev. A ~{\bf 59}, 2631 (1999).
\bibitem{Jeong02} H. Jeong and M. S. Kim, Efficient quantum computation using coherent states, Phys. Rev. A ~{\bf 65}, 042305 (2002).
\bibitem{Ralph03}  T. C. Ralph, A. Gilchrist, G. J. Milburn, W. J. Munro, and S. Glancy, Quantum computation with optical coherent states, Phys. Rev. A~{\bf 68}, 042319 (2003).
\bibitem{Lund08} A. P. Lund, T. C. Ralph, and H. L. Haselgrove, Fault-tolerant linear optical quantum computing with small-amplitude coherent states, Phys. Rev. Lett. ~{\bf 100}, 030503 (2008).
\bibitem{Myers11} C. R. Myers and T. C. Ralph, Coherent state topological cluster state production, New J. Phys. ~{\bf 13}, 115015 (2011).
\bibitem{Kim15} J. Kim, J. Lee, S.-W Ji, H. Nha, P. M. Anisimov, and J. P. Dowling, Coherent-state optical qudit cluster state generation and teleportation via homodyne detection, Opt. Comm. ~{\bf 337}, 79 (2015).
\bibitem{vanEnk01}  S. J. van Enk and O. Hirota, Entangled coherent states: Teleportation and decoherence, Phys. Rev. A ~{\bf 64}, 022313 (2001).
\bibitem{Jeong01} H. Jeong, M. S. Kim, and J. Lee, Quantum-information processing for a coherent superposition state via a mixedentangled coherent channel, Phys. Rev. A ~{\bf 64}, 052308(2001).
\bibitem{Neergaard-Nielsen13} J. S. Neergaard-Nielsen, Y. Eto, C.-W. Lee, H. Jeong, and M. Sasaki, Quantum tele-amplification with a continuous-variable superposition state, Nat. Photonics.~{\bf 7}, 439 (2013).

\bibitem{Gerry01} C. Gerry and R. Campos, Generation of maximally entangled photonic states with a quantum-optical Fredkin gate, Phys. Rev. A ~{\bf 64}, 063814 (2001).
\bibitem{Gerry02} C. Gerry, A. Benmoussa, and R. Campos, Nonlinear interferometer as a resource for maximally entangled photonic states: Application to interferometry, Phys. Rev. A ~{\bf 66}, 013804 (2002).
\bibitem{Ralph02} T. C. Ralph, Coherent superposition states as quantum rulers, Phys. Rev. A ~{\bf 65}, 042313 (2002).
\bibitem{Munro02} W. J. Munro, K. Nemoto, G. J. Milburn, and S. Braunstein, Weak-force detection with superposed coherent states, Phys. Rev. A ~{\bf 66}, 023819 (2002).
\bibitem{Campos03} R. Campos, C. Gerry, and A. Benmoussa, Optical interferometry at the Heisenberg limit with twin Fock states and parity measurements, Phys. Rev. A ~{\bf 68}, 023810 (2003).
\bibitem{Joo11} J. Joo, W. J. Munro, and T. P. Spiller, Quantum metrology with entangled coherent states, Phys. Rev. Lett. ~{\bf 107}, 083601 (2011).
\bibitem{Hirota11}  O. Hirota, K. Kato, and D. Murakami, Effectiveness of entangled coherent state in quantum metrology, http://arxiv.org/abs/1108.1517 (2011).
\bibitem{Joo12}  J. Joo, K. Park, H. Jeong, W. J. Munro, K. Nemoto, and T. P. Spiller, Quantum metrology for nonlinear phase shifts with entangled coherent states, Phys. Rev. A ~{\bf 86}, 043828 (2012).
\bibitem{Zhang13} Y. M. Zhang, X. W. Li, W. Yang, and G. R. Jin, Quantum Fisher information of entangled coherent states in the presence of photon loss, Phys. Rev. A ~{\bf 88}, 043832 (2013).

\bibitem{Vasconcelos10} H. M. Vasconcelos, L. Sanz, S. Glancy, All-optical generation of states for “Encoding a qubit in an oscillator”, Opt. Lett. 35, 3261–3263 (2010).
\bibitem{Weigand18} D. J. Weigand, B. M. Terhal, Generating grid states from Schrödinger-cat states without postselection, Phys. Rev. A 97, 022341 (2018).
\bibitem{Shunya24} S. Konno {\bf et al.}, Logical states for fault-tolerant quantum computation with propagating light, Science {\bf 383}, 289 (2024).

\bibitem{Li17} L. Li, C.-L. Zou, V. V. Albert, S. Muralidharan, S. M. Girvin and L. Jiang, Cat codes with optimal decoherence suppression for a lossy bosonic channel, Phys. Rev. Lett. ~{\bf 119}, 030502 (2017).
\bibitem{Hastrup22} J. Hastrup, U. L. Andersen, All-optical cat-code quantum error correction, Phys. Rev. Res. 4, 043065. (2022).


\bibitem{Wilson02} D. Wilson, H. Jeong and M. S. Kim, Quantum nonlocality for a mixed entangled coherent state, J. Mod. Opt. {\bf 49}, 851 (2002).
\bibitem{Park15}  C.-Y. Park and H. Jeong, Bell-inequality tests using asymmetric entangled coherent states in asymmetric lossy environments, Phys. Rev. A {\bf 91}, 042328 (2015).
 
\bibitem{Paternostro10} M. Paternostro and H. Jeong, Testing non-local realism with entangled coherent states, Phys. Rev. A. {\bf 81}, 032115 (2010). 
\bibitem{Lee11} C.-W. Lee, M. Paternostro, and H. Jeong, Faithful test of nonlocal realism with entangled coherent states, Phys. Rev. A {\bf 83}, 022102 (2011).

\bibitem{Neergaard-Nielsen06} J. S. Neergaard-Nielsen, B. M. Nielson, C. Hettich, K. M{\o}lmer, and E. S. Polzik, Generation of a Superposition of Odd Photon Number States for Quantum Information Networks, Phys. Rev. Lett.~{\bf 97}, 083604 (2006).
\bibitem{Ourjoumtsev07} A. Ourjoumtsev, H. Jeong, R. Tualle-Brouri and Ph. Grangier, Generation of optical ‘Schrödinger cats’ from photon number states, Nature. ~{\bf 448}, 784 (2007).
\bibitem{Wakui07} K. Wakui, H. Takahashi, A. Furusawa, and M. Sasaki, Photon subtracted squeezed states generated with periodically poled KTiOPO4, Opt. Express~{\bf 15}, 3568 (2007).
\bibitem{Ourjoumtsev09}  A. Ourjoumtsev, F. Ferreyrol, R. Tualle-Brouri, and P. Grangier, Preparation of non-local superpositions of quasi-classical light states, Nature Phys. ~{\bf 5}, 189–192 (2009).
\bibitem{Ourjoumtsev06} A. Ourjoumtsev, R. Tualle-Brouri, J. Laurat, and P. Grangier, Generating optical Schrodinger kittens for quantum information processing, Science ~{\bf 312}, 83 (2006).


\bibitem{Jeong05}  H. Jeong, A. P. Lund, and T. C. Ralph, Production of superpositions of coherent states in traveling optical fields with inefficient photon detection, Phys. Rev. A ~{\bf 72}, 013801 (2005).

\bibitem{Lund04} A. P. Lund, H. Jeong, T. C. Ralph, and M. S. Kim, Conditional production of superpositions of coherent states with inefficient photon detection, Phys. Rev. A ~{\bf 70}, 020101 (2004).




\bibitem{Lvovsky2017}
D.V. Sychev, A.E. Ulanov, A.A. Pushkina, M.W. Richards, I.A. Fedorov, and A.I. Lvovsky, Enlargement of optical Schrödinger's cat states, Nature Photonics {\bf 11}, 379 (2017).

\bibitem{Oh18} C. Oh and H. Jeong, Efficient amplification of superpositions of coherent states using input states with different parities, J. Opt. Soc. Am. B ~{\bf 35}, 2933 (2018).

\bibitem{Lag13} A. Laghaout, J.S. Neergaard-Nielsen, I. Rigas, C. Kragh, A. Tipsmark, and U.L. Andersen, Amplification of realistic Schrödinger-cat-state-like states by homodyne heralding, Phys. Rev. A {\bf 87}, 043826 (2013).



\bibitem{Sheng13} Y.-B. Sheng, L. Zhou, and G.-L Long, Hybrid entanglement purification for quantum repeaters, Phys. Rev. A ~{\bf 88}, 022302 (2013).
\bibitem{Caves88} C. Caves, Quantum limits on noise in linear amplifiers, Phys. Rev. D~{\bf 26}, 1817 (1982).
\bibitem{Parigi07} V. Parigi, A. Zavatta, and M. Bellini, Manipulating thermal light states by the controlled addition and subtraction of single photons, Laser Phys. Lett.~{\bf 5}, 3 (2007).
\bibitem{Zavatta11} A. Zavatta, J. Fiur\'a\v sek and M. Bellini, A high-fidelity noiseless amplifier for quantum light states, Nat. Photonics ~{\bf 5}, 52 (2011).

\bibitem{Park16} J. Park, J. Joo, A. Zavatta, M. Bellini and H. Jeong, Efficient noiseless linear amplification for light fields with larger amplitudes, Opt. Express ~{\bf 24}, 1331 (2016).

\bibitem{Fadrny24} J. Fadrn\'y, M. Neset, M. Bielak, Miroslav Je\v zek, Jan B\'ilek, and J. Fiur\'a\v sek, Experimental preparation of multiphoton-added coherent states of light. npj Quantum Inf ~{\bf 10}, 89 (2024)



\bibitem{Helstrom76} C. W. Helstrom, Mathematics in Science and Engineering (Academic Press, New York, 1976), Vol. 123.
\bibitem{Braunstein94} S. L. Braunstein and C. M. Caves, Statistical distance and the geometry of quantum states, Phys. Rev. Lett. ~{\bf 72}, 3439 (1994).
\bibitem{Barndorff-Nielsen00} O. E. Barndorff-Nielsen, R. D. Gill, Fisher information in quantum statistics, J. Phys. A~{\bf 30}, 4481 (2000).
\bibitem{Paris09} M. G. A. Paris, Quantum estimation for quantum technology, Int. J. Quantum Inf. ~{\bf 7}, 125 (2009).
\bibitem{Petz10} D. Petz, C. Ghinea, Introduction to Quantum Fisher information, In Quantum Probability and Related Topic, 261 (2011)

\bibitem{Usuga10} M. A. Usuga, C. R. M\" uller, C. Wittmann, P. Marek, R. Filip, C. Marquardt, G. Leuchs, and U. L. Andersen, Noise-powered probabilistic concentration of phase information, Nat. Phys.~{\bf 6}, 767 (2010).
\bibitem{Genoni11} M. G. Genoni, S. Olivares, and M. G. A. Paris, Optical Phase Estimation in the Presence of Phase Diffusion, Phys. Rev. Lett.~{\bf 106}, 153603 (2011).
\bibitem{Berni15} A. A. Berni, T. Gehring, B. M. Nielsen, V. H\" andchen, M. G. A. Paris, and U. L. Andersen, Ab initio quantum-enhanced optical phase estimation using real-time feedback control, Nat. Photonics ~{\bf 9}, 577 (2015).
\bibitem{Izumi16} S. Izumi, M. Takeoka, K. Wakui, M. Fujiwara, K. Ema, and M. Sasaki, Optical phase estimation via the coherent state and displaced-photon counting, Phys. Rev. A~{\bf 94}, 033842 (2016).
\bibitem{Grangier92} Ph. Grangier, J.-M. Courty, and S. Reynaud, Characterization of nonideal quantum non-demolition measurements, Opt. Commun.~{\bf 89}, 99 (1992).

\bibitem{note1} In a real experiment, a single-photon addition is typically heralded by detecting a photon such as in Ref.~\cite{Z2004}. However, we consider a simpler conceptual scheme heralded by the vacuum measurement which is sufficient for our purpose.

\bibitem{Z2004} A. Zavatta, S. Viciani, and M. Bellini, 
Quantum-to-classical transition with single-photon-added coherent states of light,
Science {\bf 306}, 660 (2004).


\end{thebibliography}
\end{document}